\documentclass{article}
\usepackage{amssymb,hhline}

 \def\rql{\mbox{\strut\smash{\raisebox{-1.4ex}{''}}\hspace{-0.05em}}}
 \def\rqr{\mbox{\hspace{0em}``}}

\newcommand*{\mytheorem}{\mbox{T~h~e~o~r~e~m.~}}

\makeatletter
\renewcommand{\@oddfoot}{--- ~August, 2006~ ---\hfil
\raisebox{0.3ex}{\tiny --- ~D.~A.~Arbatsky~ ``The certainty
principle (review)''~ ---}\hfil --- ~\thepage~ ---}
 \makeatother

\newcommand{\titu}{%
\begin{tabular}{c}
 {\bf uncertainty principle}\\
 \footnotesize (Heisenberg, 1927)
\end{tabular}}

\newcommand{\titc}{%
\begin{tabular}{c}
 {\bf certainty principle}\\
 \footnotesize (Arbatsky, 2005)
\end{tabular}}

\newcommand{\xpuk}{%
\begin{tabular}{c}
 $ \Delta_\rangle X\,\Delta_\rangle P \,\geqslant\, \frac\hbar2 $\\
 \footnotesize (Kennard, 1927)
\end{tabular}}

\newcommand{\xpua}{%
\begin{tabular}{c}
 $ \delta_\rangle X\,\Delta_\rangle P \,\geqslant \hbar $\\
 \footnotesize (Arbatsky, 2006)
\end{tabular}}

\newcommand{\xpco}{%
\begin{tabular}{c}
 $ |\delta x|\, \Delta_\rangle P \geqslant \hbar $\\
 \footnotesize (Arbatsky, 2005)
\end{tabular}}

\newcommand{\xpca}{%
\begin{tabular}{c}
 $\Delta_\rangle (\,
   \delta x_1\,P_1\, +\,
   \delta x_2\,P_2\, +\,
   \delta x_3\,P_3\,
   ) \geqslant \hbar $\\
 \footnotesize (Arbatsky, 2005)
\end{tabular}}

\newcommand{\pjuj}{%
\begin{tabular}{c}
 $\displaystyle
 \Delta'_\rangle\Phi \, \Delta_\rangle J \,\geqslant\,
 \frac\hbar2
 \,
 \left[\,
  1\, -\, \frac3{\pi^2}\, (\Delta'_\rangle\Phi)^2
 \,\right]$\\
 \footnotesize (Judge, 1964)
\end{tabular}}

\newcommand{\pjua}{%
\begin{tabular}{c}
 $\delta_\rangle\Phi \ \geqslant\
 \min\ (\ \hbar\,/\,\Delta_\rangle J\ ;\ \pi\ )$\\
 \footnotesize (Arbatsky, 2006)
\end{tabular}}

\newcommand{\pjco}{%
\begin{tabular}{c}
 $ |\delta\varphi|\, \Delta_\rangle J \geqslant \hbar $\\
 \footnotesize (Arbatsky, 2005)
\end{tabular}}

\newcommand{\pjca}{%
\begin{tabular}{c}
 $\Delta_\rangle (\,
   \delta\varphi_1\,J_1\, +\,
   \delta\varphi_2\,J_2\, +\,
   \delta\varphi_3\,J_3\,
   ) \geqslant \hbar$\\
 \footnotesize (Arbatsky, 2005)
\end{tabular}}

\newcommand{\tecmt}{%
\begin{tabular}{c}
 $|\delta t|\, \Delta_\rangle H \geqslant \hbar$\\
 \footnotesize (Mandelshtam, Tamm, 1945)
\end{tabular}}

\newcommand{\alca}{%
\begin{tabular}{c}
 $\Delta_\rangle\,
   (\,-\,\delta x_\mu\, P_\mu\, +
      \frac12\,
      \delta\omega_{\mu\nu}\, J_{\mu\nu} \,) \geqslant \hbar$\\
\footnotesize (Arbatsky, 2005)
\end{tabular}}

\newcommand{\tcptab}{%
\begin{tabular}{||c||c|c||}
 \hhline{|t:=:t:=t=:t|}
 \hspace{2cm} & \hspace{7cm} & \hspace{7cm} \\
  & \multicolumn{2}{c||}{
    \titu\hspace{12.5mm}
    $\Longleftarrow$
    \hspace{14.5mm}\titc\hspace{3mm} } \\
 &&\\
 \hhline{|:=::=:=:|}
 &&\\
 & \xpuk & \xpco \\
 $x$, $p$ & & \\
 & \xpua & \xpca \\
 &&\\
 \hhline{||-||-|-||}
 &&\\
 & \pjuj & \pjco \\
 $\varphi$, $J$ & & \\
 & \pjua & \pjca \\
 &&\\
 \hhline{||-||-|-||}
 &&\\
 $t$, $E$ & --- & \tecmt \\
 &&\\
 \hhline{||-||-|-||}
 &&\\
 all & --- & \alca \\
 &&\\
 \hhline{|b:=:b:=t=:b|}
\end{tabular}}

\begin{document}


\title{The certainty principle (review)}
\author{D.~A.~Arbatsky\footnote{ http://daarb.narod.ru/ , http://wave.front.ru/ }}
\date{August, 2006}
\maketitle

\begin{abstract}
The certainty principle (2005) allowed to conceptualize from the
more fundamental grounds both the Heisenberg uncertainty principle
(1927) and the Mandelshtam-Tamm relation (1945). In this review I
give detailed explanation and discussion of the certainty
principle, oriented to all physicists, both theorists and
experimenters.
\end{abstract}

\section*{Historical comments}

\paragraph{Uncertainty principle.}
A reader interested in the history of the {\it uncertainty}
principle should read, for example, the review \cite{HilgUff2001}.
Here I give only some notes, that are important for the following
recital.

The uncertainty principle was suggested by Heisenberg in 1927
\cite{Heis1927}. Heisenberg formulated it as follows:
\begin{itemize}
\item
The more precisely the position is determined, the less precisely
the momentum is known in this instant, and vice versa.
\end{itemize}
For uncertainties of coordinate and momentum the following
relation was suggested:
\begin{equation}
 \Delta_\rangle X\,\Delta_\rangle P \,\sim\, \hbar \ .
 \label{HeisRel1}
\end{equation}

Considering concrete examples, Heisenberg gave just qualitative
formulation. But for the notion of ``uncertainty'' he did not give
exact definition.

Soon Kennard \cite{Kennard1927} gave exact mathematical
formulation for the case of coordinate and momentum. Assuming
``commutation relation'' $[X,P]=i\hbar$, he has shown that for an
arbitrary quantum state $\rangle$ the following relation takes
place:
\begin{equation}
 \Delta_\rangle X\,\Delta_\rangle P \,\geqslant\,
 \frac\hbar2 \ .
 \label{KennardRel}
\end{equation}
And here
\[
 \Delta_\rangle X =
   \left\langle\bigl(\,
    X-\langle X\rangle
   \,\bigr)^2\right\rangle^{1/2}
 \ , \qquad
 \Delta_\rangle P =
   \left\langle\bigl(\,
    P-\langle P\rangle
   \,\bigr)^2\right\rangle^{1/2}
 \ .
\]
It means that, in accordance with probabilistic interpretation of
quantum mechanics, it was suggested to understand the
uncertainties of coordinate and momentum as standard deviations of
these observables.

Because of simplicity of the proof, the relation
(\ref{KennardRel}) became conventional mathematical expression of
the uncertainty principle and appeared in all textbooks on quantum
mechanics. All criticism (including that about correspondence to
practical experiment) \cite{HilgUff2001} was mainly ignored. But
soon we will see that the certainty principle allows to get
alternate inequalities, describing the uncertainty principle,
which turn out to be more sapid from practical point of view.

Later the mathematical proof was generalized in many ways. And it
was extended to other pairs of non-commuting observables. So, in
order to avoid ambiguity in the term {\it uncertainty principle},
let us give here some other, more up-to-date and general,
formulation of this principle\footnote{Such a formulation not only
allows to better understand, what exactly is ``uncertain'', but
also allows to clearly see the contrast with the {\it certainty}
principle.}:
\begin{itemize}
\item
If one tries to describe the dynamical state of a {\it quantum}
particle by methods of {\it classical} mechanics, then precision
of such description is limited in principle. The {\it classical}
state of the particle turns out to be {\it badly} defined. This
uncertainty can be mathematically expressed by different
inequalities, describing spreads of values of observables that
have semiclassical limit.
\end{itemize}

Of course, Heisenberg could not formulate the uncertainty
principle in this form. It would contradict the logic of
historical moment. In 1927 physics came from classical mechanics
to quantum, and the uncertainty principle was considered as a way
from the old theory to new.

But today, when quantum mechanics is already well-established, the
uncertainty principle is just a method of qualitative estimation
of precision of semiclassical approximation. And there are no
reasons to think that this principle is more fundamental than
mathematical formalism of quantum mechanics.


\paragraph{Mandelshtam-Tamm relation.}
So far as for coordinate and momentum there is an uncertainty
relation like (\ref{HeisRel1}), general arguments, connected with
the theory of relativity,\footnote{But soon we will see that these
arguments are {\it incorrect}, because existence of the
uncertainty principle is connected with specific peculiarities of
{\it non-relativistic} approximation.} point out that, seemingly,
an uncertainty relation for time and energy like the following
should exist:
\begin{equation}
 \Delta_\rangle T\,\Delta_\rangle H \,\sim\, \hbar \ .
 \label{TH1}
\end{equation}
Heisenberg himself suggested such a relation. But even in that
time it was clear \cite{Pauli1933} that explanation of such a
relation must be more difficult, because there is no such a
quantum-mechanical observable as ``time'', and it is unclear what
should be understood as ``uncertainty'' of time.

Nevertheless, physicists wanted to believe that the uncertainty
principle was a {\it fundamental} physical principle, and attempts
to ground relation (\ref{TH1}) did not stop up to now. These
attempts mainly tried to show that the notion ``uncertainty of
time'' was meaningful. And it was usually made by analysis of
measurement process, i.~e. to analysis of interaction of quantum
particle with apparatus like Heisenberg microscope. (a detailed
review with critical analysis see in \cite{Voron1981})

I cannot say that those attempts were unsuccessful. Nevertheless,
I think that such an approach does not correspond to general
methodology of quantum mechanics. The matter is that analytical
formalism of quantum mechanics is oriented to studying quantum
systems as independent physical objects. This analytical formalism
does not study details of interaction of apparatuses with the
quantum system.

In 1945 Mandelshtam and Tamm \cite{MandTamm1945} gave rigorous
mathematical formulation  of relation\footnote{Here it is written
a little differently, so that connection with further discussion
would be clearer.}
\begin{equation}
 |\delta t|\, \Delta_\rangle H \geqslant \hbar \ .
 \label{MandTammRel}
\end{equation}
Mandelshtam and Tamm named this relation ``the uncertainty
relation between energy and time in nonrelativistic quantum
mechanics'', because, obviously, they thought that they found
mathematical expression of the uncertainty principle for energy
and time. Correspondingly, further attempts to ground {\it
uncertainty} relation for energy and time were generally attempts
to show that the quantity $\delta t$ in Mandelshtam-Tamm relation
can be understood in some sense as ``uncertainty'' of time.

But, I believe that there are no reasons convincing enough to call
Mandelshtam-Tamm relation as {\it uncertainty} relation. As it was
explained above, the uncertainty principle is not more fundamental
than mathematical formalism of quantum mechanics itself. And what
is more, analysis of relativistic quantum systems shows that this
principle is {\it less} fundamental. Therefore there are no
reasons to think that fundamental uncertainty relation like
(\ref{TH1}) must exist.

Soon we will see that Mandelshtam-Tamm relation in the case of
closed systems is a consequence of other physical principle, the
{\it certainty} principle, which is more fundamental than the
Heisenberg uncertainty principle. And the quantity $\delta t$ in
the inequality (\ref{MandTammRel}), if we make transformation from
Schr\"odinger to Heisenberg representation (or to
RCQ-representation), is a logarithmic coordinate on Poincare
group. For some purposes this coordinate can be interpreted as
``uncertainty of time'', but, generally speaking, there is no need
in such interpretation.

As regards non-closed systems, with arbitrary dependence of
Hamiltonian on time, Mandelshtam-Tamm relation should be
considered as relation for the speed of quantum evolution. And it
is quite independent relation, not mathematically equivalent to
the certainty relation for time and energy.


\paragraph{Certainty principle.}

When we raise a question about application of ideas connected with
the theory of relativity in quantum mechanics, it is worth to
think about what we know about relativistic invariance of quantum
theory at all.

The group of invariance of Minkowski space\footnote{Of
four-dimensional space-time of the special theory of relativity.}
is Poincare group. With invariance with respect to this group of
laws of nature, in general, and of quantum mechanics, in
particular, is connected the notion of relativistic invariance.

All known physically important quantum systems, to which action of
Poincare group can be applied, are quantum fields (or their
subsystems). And before 2002 mathematical construction of quantum
fields was made by such recipes\footnote{Recipes, because formal
mathematical process did not exist.}, which made action of
Poincare group on space of states of quantum field hidden. In
fact, physicists were satisfied by statement that Poincare group
acts on space of states not-explicitly, by some quite complicated
formulas.

When relativistic canonical quantization (RCQ) appeared
\cite{Arb2002} (for introduction review \cite{Arb2005wircq} is
recommended) it became clear how action of Poincare group is
transferred from Minkowski space to the space of states of quantum
field.\footnote{Specific feature of RCQ method is construction of
quantum fields in the frame of four-dimensional geometry of
Minkowski space, without artificial separation of space and time.}
It became also obvious the following:
\begin{itemize}
\item
The notion of coordinate as quantum-mechanical observable for
relativistic systems is not natural at all, not even
non-fundamental. (Attempts in the past to introduce coordinates
for some systems, for example \cite{NewtWig1949}, seem to be quite
artificial.)
\item
For this reason the uncertainty principle {\it can not} be
fundamental, in the sense of its generalization for relativistic
quantum theory.
\end{itemize}

Using methodology of RCQ, I have formulated \cite{Arb2005tcp} the
{\it certainty} principle:
\begin{itemize}
\item
If one describes the dynamical state of a quantum particle
(system) by methods of {\it quantum} mechanics, then the {\it
quantum} state of the particle (system) turns out to be {\it well}
defined. This certainty of the quantum dynamical state means that
``small'' space-time transformations can not substantially change
the quantum state. And for the case of Poincare group, for
transformations, that can substantially change the quantum state,
we have estimation\footnote{Here and later we imply relativistic
summation by pairs of the same Greek indices:
\[
 a_\mu\, b_\mu = a_0\, b_0 - a_1\, b_1 - a_2\, b_2 - a_3\, b_3 \ .
\].}:
\begin{equation}
 \textstyle
 \Delta_\rangle\,
   (\,-\,\delta x_\mu\, P_\mu\, +
      \frac12\,
      \delta\omega_{\mu\nu}\, J_{\mu\nu} \,) \geqslant \hbar \ ,
 \label{CertPoin}
\end{equation}
Here $P_\mu$ is a vector operator of energy-momentum, $J_{\mu\nu}$
is a tensor operator of the four-dimensional angular momentum,
$\delta x_\mu$ and $\delta\omega_{\mu\nu}$ are the standard
logarithmic coordinates of the Poincare group.
\end{itemize}

So far as the certainty principle is universal, it works well also
in usual quantum mechanics, including {\it non-relativistic}. So,
it seems to be surprising that it was not formulated earlier.
Nevertheless, there are two explanations for this.

First, this principle is formulated naturally with basic notions
of RCQ. In particular, it is most natural in formulation of the
certainty principle to understand the notion of {\it dynamical
state} within RCQ interpretation. This difficulty, seemingly, was
the main.

Second, in order to come to the certainty principle from the usual
quantum mechanics, it is necessary to use the theory of quantum
angle (Fubini-Study metric). But existence of this metric, though
not difficult to prove, is not obvious. This metric was discovered
only in 1905 by Fubini \cite{Fubini1904} and Study
\cite{Study1905}, and most physicists do not know about it.

\section*{Mathematical formalism}

\paragraph{Quantum angle (Fubini-Study metric).}
Consider the unit sphere $\|x\|=1$ in the usual three-dimensional
Euclidean space $\mathbb R^3$. The distance measured on the
surface of the sphere between any two points is equal to the angle
between rays starting from the center of coordinates and coming
through these points.

So, the connection of angles with rays on the sphere makes obvious
the important property of angle: it satisfies the ``triangle
inequality'':
\begin{equation}
 \angle(a,c) \leqslant \angle(a,b) + \angle(b,c) \ .
 \label{TrianIneq}
\end{equation}

It is known, that angle in this example is expressed through the
scalar product $\langle\cdot|\cdot\rangle$ in the space
 $\mathbb R^3$ by formula:
\[
 \angle(a,b)\, = \,\arccos\, \langle a|b\rangle \ .
\]
Here in the right part of the formula vectors, whose ends are in
the points of intersections of the rays $a$ and $b$ with the unit
sphere, are denoted by the same letters as rays. Possible values
of angle in this example are in the range from $0$ to $\pi$.

Let us now consider angles not between rays, but between lines
coming through the center of coordinates. In this case possible
values of angle are in the more narrow range from $0$ to $\pi/2$.
The formula for the angle in this case is slightly changed:
\begin{equation}
 \angle(a,b)\, = \,\arccos\,\bigl| \langle a|b\rangle \bigr| \ .
 \label{QAngle}
\end{equation}

Will the angle satisfy the triangle inequality (\ref{TrianIneq})
in this case? The answer to this question does not seem to be
immediately obvious, because for every line we have two points on
the sphere. But elementary logical reasoning allows to reduce this
example to the previous. The answer turns out to be positive.

Consider now the case of the three-dimensional complex Hilbert
space $\mathbb C^3$. Instead of real lines in this case we
obviously have complex lines, i.~e. one-dimensional complex
subspaces. Every such a subspace intersects with the unit sphere
even not in two, but in infinite number of points. These points
are different from each other by complex factor, whose absolute
value is equal to one.

The angle in this case is defined by the same formula
(\ref{QAngle}), but the scalar product here corresponds to the
space $\mathbb C^3$. Possible values of angle, obviously, also are
in the range from $0$ to $\pi/2$.

Will angle in this case satisfy the triangle inequality
(\ref{TrianIneq})? The answer in this case turns out to be
positive also. This fact was proved by Fubini \cite{Fubini1904}
and Study \cite{Study1905} (quite simple proof is also given in
\cite{Arb2005tcp}), so the angle between complex lines in Hilbert
space is called {\it Fubini-Study metric}.

Of course, the fact that triangle inequality takes place cannot
depend on the dimensionality of the Hilbert space, because when we
check the triangle inequality we always can consider the
corresponding three-dimensional subspace.

So far as in quantum mechanics the space of states of a quantum
system $\mathcal H$ is complex Hilbert space, Fubini-Study metric
can be naturally considered as a measure of difference of quantum
states. In this context we will call this metric {\it quantum
angle}, because it allows to form natural collocations like
``quantum angular speed''.


\paragraph{Quantum angular speed.}
Let now the vector $r$ depend on the real parameter $t$:
$t\in\mathbb R$, $r(t)\in\mathcal H$, $\|\,r(t)\,\|=1$.

Let us define {\it the quantum velocity} $v(t)$ by the formula:
\[
 v(t) = \dot r(t) =
    \lim_{\delta t\to0}
    \frac{r(t+\delta t)-r(t)}{\delta t} \ .
\]

Let us define also {\it the quantum angular speed} $\omega(t)$ by
the formula:
\[
 \omega(t) = \lim_{\delta t\to0}
     \frac{\angle\,\big(\,r(t+\delta t)\,,\,r(t)\,\big)}{|\,\delta t\,|} \ .
\]

In order to express $\omega(t)$ through $v(t)$ let us decompose
$v(t)$ into the two orthogonal components:
\[
 v_\parallel(t) = r(t)\,\langle r(t)|v(t)\rangle \ , \qquad
 v_\perp(t) = v(t) - v_\parallel(t) \ .
\]

This decomposition formally looks the same as the corresponding
decomposition in the real case, which is well known in kinematics
of point. But we should not rely on this analogy too much. The
matter is that in the real case existence of the parallel
component of velocity is necessarily connected with growth of the
norm of the vector. But in the complex case under consideration
vector is supposed to have always unit norm. Existence of parallel
component of velocity turns out connected with the possibility of
multiplication of the vector by complex phase factor, that is
equal to one. Nevertheless, the following theorem is true

\mytheorem{}
 {\it The quantum angular speed is equal to the norm of the orthogonal
component of the quantum velocity:
\begin{equation}
 \omega(t)=\|\,v_\perp(t)\,\| \ .
 \label{QAngSpeedQVel}
\end{equation}}

Before proving this theorem let us note that in the real case
equality (\ref{QAngSpeedQVel}) is {\it definition} of angular
speed $\omega$. And angle in the real case is {\it defined} as
integral of angular speed along extreme arc. After that
trigonometrical function $\arccos$ is {\it defined} for {\it
finite} angles.

Straightforward transfer of such logic to the case of complex
Hilbert space is problematic, because there are always many
extreme arcs, and none of them is so simple that appearance of
$\arccos$ function would be geometrically obvious.

So, here we need to study directly infinitesimal  angles, i.~e.
work around singular point of $\arccos$ function. In order to
escape from this, let us use Parseval equality to change $\arccos$
to $\arcsin$:
\[
 \angle\,\big(\,r(t+\delta t)\,,\,r(t)\,\big) \,=\,
 \arccos\,\big|\,\langle\,r(t+\delta t)\,|\,r(t)\,\rangle\,\big| \,=\,
\]
\[
 \,=\,
 \arcsin\,
   \big\|\,
   r(t+\delta t) -
   r(t) \,\langle\, r(t) \,|\, r(t+\delta t) \,\rangle
   \,\big\| \,=\,
\]
\[
 \,=\,
   \arcsin\,
    \big\|\,
    r(t+\delta t) \,-\, r(t) \,-\,
    r(t) \,
     \langle\,
        r(t) \,|\, r(t+\delta t)-r(t)
     \,\rangle
    \,\big\|
 \,=\,
\]
\[
 \,=\,
   \arcsin\,
    \big\|\,
    v(t)\,\delta t \,+\, o(\delta t) \,-\,
    r(t) \,
     \langle\,
        r(t) \,|\, v(t)\,\delta t + o(\delta t)
     \,\rangle
    \,\big\|
 \,=\,
\]
\[
 \,=\,
   \arcsin\,
    \Big\|\,
     \big(v(t)\,-\,
     r(t) \,
     \langle\,
        r(t) \,|\, v(t)
     \,\rangle\big)
     \,\delta t \,+\, o(\delta t)
    \,\Big\|
 \,=\,
\]
\[
 \,=\,
   \arcsin\,
    \big\|\,
    v_\perp(t)\,\delta t \,+\, o(\delta t)
    \,\big\|
 \,=\,
   \arcsin\,
    \big(\,
    \|v_\perp(t)\|\,|\delta t| \,+\, o(\delta t)
    \,\big)
 \,=\,
\]
\[
 \,=\,
 \|v_\perp(t)\|\,|\delta t| \,+\, o(\delta t) \ .
\]
Dividing by $|\delta t|$, we get (\ref{QAngSpeedQVel}).
$\blacksquare$


\paragraph{Continuous unitary groups.}
Suppose now that we have in the space of states of quantum system
a self-adjoint operator $A=A^*$. The set of unitary operators of
type $U(\delta s)=e^{-\,i\,\delta s\,A/\hbar}$ (where $\delta s$
runs the set of all real numbers, and $\hbar$ is fixed positive
real number (in quantum mechanics this is Planck's constant))
forms strongly continuous one-parameter unitary group:
 $U(\delta s_1)\,U(\delta s_2)\,=\, U(\delta s_1 + \delta s_2)$.
The operator $A$ here is called {\it generator} of this group.

And suppose that the state vector is changed by the action of this
group:
\[
 r(\delta s) \,=\, |\,\delta s\,\rangle \,=\,
 U(\delta s)\,\rangle \,=\,
 e^{-\,i\,\delta s\,A/\hbar} \,\rangle \ .
\]
Here $|\,\delta s\,\rangle\in\mathcal H$ is another notation of
the vector with parameter equal to $\delta s$;
 $\rangle\in\mathcal H$ is a fixed ket-vector.

Suppose now that the function $r(\delta s)$ is differentiable.
Then the quantum velocity is expressed by the formula:
\[
 \textstyle
 v(\delta s) \,=\,
 \frac1{i\hbar}\, A \, e^{-\,i\,\delta s\,A/\hbar} \,\rangle \,=\,
 \frac1{i\hbar}\, A \, |\,\delta s\,\rangle \ .
\]

The mean of the operator $A$ does not depend on time:
\[
 \overline A \,=\,
 \langle\,\delta s\,|\, A \, |\,\delta s\,\rangle \,=\,
 \langle\,e^{+\,i\,\delta s\,A/\hbar} \, A \,
    e^{-\,i\,\delta s\,A/\hbar} \,\rangle \,=\,
 \langle\, A \, e^{+\,i\,\delta s\,A/\hbar} \,
    e^{-\,i\,\delta s\,A/\hbar} \,\rangle \,=\,
 \langle\, A \, \rangle \ .
\]

Therefore the components of the quantum velocity can be written
just as:
\[
 \textstyle
 v_\parallel(\delta s) \,=\,
 |\,\delta s\,\rangle \, \langle\,\delta s\,|\,
 \frac1{i\hbar}\, A \, |\,\delta s\,\rangle \,=\,
 \frac1{i\hbar}\,\overline A \, |\,\delta s\,\rangle
\]
\[
 \textstyle
 v_\perp(\delta s) \,=\,
 \frac1{i\hbar}\,\left(A-\overline A\right) \,
 |\,\delta s\,\rangle \ .
\]

The quantum angular speed turns out to be independent of the
parameter also:
\[
 \textstyle
 \omega(\delta s)
 \,=\,
 \|\,v_\perp(\delta s)\,\|
 \,=\,
 \frac1\hbar\,
 \langle\,\delta s\,|\,
    \left(A-\overline A\right)^2
 \, |\,\delta s\,\rangle^{1/2}
 \,=\,
 \frac1\hbar\,
 \langle\,
    e^{+\,i\,\delta s\,A/\hbar} \,
    \left(A-\overline A\right)^2 \,
    e^{-\,i\,\delta s\,A/\hbar}
 \,\rangle^{1/2}
 \,=\,
\]
\[
 \textstyle
 \,=\,
 \frac1\hbar\,
 \langle\,
    \left(A-\overline A\right)^2 \,
    e^{+\,i\,\delta s\,A/\hbar} \,
    e^{-\,i\,\delta s\,A/\hbar}
 \,\rangle^{1/2}
 \,=\,
 \frac1\hbar\,
 \langle\,
    \left(A-\overline A\right)^2
 \,\rangle^{1/2}
 \,=\,
 \frac1\hbar\, \Delta_\rangle A  \ .
\]
Here $\Delta_\rangle A$ is a short notation for the standard
deviation of $A$ in the state $\rangle$.

So, we have proved the important

\mytheorem{}
 {\it Standard deviation of the generator of one-parameter unitary
group is equal to the quantum angular speed, multiplied by
Planck's constant $\hbar$. And both quantities remain conserved
under the action of the group.}

Determine now, how different can become the initial $\rangle$ and
final $|\,\delta s\,\rangle$ vectors. The triangle inequality, in
usual way, is generalized to arbitrary curves \cite{Arb2005tcp}.
Therefore, the angle between initial and final vectors cannot be
greater than increment of parameter, multiplied by quantum angular
speed:
\begin{equation}
 \textstyle
 \angle\,\big(\,|\,\delta s\,\rangle\,,\,\rangle\,\big) \,\leqslant\,
 |\delta s| \, \frac1\hbar\, \Delta_\rangle A  \ .
 \label{AngEst}
\end{equation}

For physical applications it is convenient to introduce some
qualitative border, when two state vectors can be considered
different {\it substantially}. So, we will say that two vectors
are different {\it substantially}, if the angle between them is
greater or equal to $1$.

Let us reformulate inequality (\ref{AngEst}) as the following
theorem.

\mytheorem{} (certainty principle)
 {\it So that under the action of strongly continuous one-parameter
unitary group $U(\delta s)=e^{-\,i\,\delta s\,A/\hbar}$ the
initial state vector $\rangle$ changes substantially, it is
necessary to satisfy the inequality:
\begin{equation}
 |\delta s|\, \Delta_\rangle A \geqslant \hbar \ .
 \label{Certainty1}
\end{equation}}

This theorem is easily generalized to the case when the group is
many-parameter. Namely, if $\delta s$ and $A$ have matrix indices,
implying summation, then the inequality (\ref{Certainty1}) can be
re-written for this case a little differently:
\[
 \Delta_\rangle (\,\delta s_j\, A_j\,) \geqslant \hbar \ .
\]
And here $\delta s_j$ are the so called logarithmic coordinates of
the group.

It is worth also to mention the following. When proving the
theorem we supposed that quantum velocity is well-defined, i.~e.
the corresponding derivative of the vector with respect to the
parameter exists and belongs to the Hilbert space. But, if the
generator is an unbounded operator, for some state vectors it can
be not so. Nevertheless, the statement of the theorem turns out to
be true even in this case, because in this case the value
$\Delta_\rangle A$ turns out to be infinite. Therefore, when
parameter is not equal to zero, in the left part of the inequality
we have infinity (which is, of course, greater than Planck's
constant).

So, the statement of the theorem turns out to be true for {\it
any} state vector $\rangle$ and {\it any} (self-adjoint) generator
$A$.

\section*{Subgroups of the Poincare group}

The theorem presented at the end of the previous section is the
most general formulation of the certainty principle. But in
different situations the unitary group under consideration can be
representation of some concrete physical symmetry group. The most
important example is Poincare group, for which the general
certainty relation (\ref{CertPoin}) was already presented above.
This group has some physically important subgroups, which are so
practically important, that we will discuss them here in more
detail.


\paragraph{Space translations.}
It is known that generators of three parameter group of space
translations are components of the vector operator of momentum
$P_1$, $P_2$ and $P_3$. Consequently, the certainty relation is
written as
\begin{equation}
 \Delta_\rangle (\,
   \delta x_1\,P_1\, +\,
   \delta x_2\,P_2\, +\,
   \delta x_3\,P_3\,
 ) \geqslant \hbar  \ .
 \label{CertTrans}
\end{equation}
Here numbers $\delta x_1$, $\delta x_2$ and $\delta x_3$ form
components of the vector that defines the translation.

If we suppose now that $\delta x_2=\delta x_3=0$, and also omit
the index, then the inequality (\ref{CertTrans}) can be written
as:
\[
 |\delta x|\, \Delta_\rangle P \geqslant \hbar \ .
\]

Suppose now, that for the system under consideration we were able
to find  some observable $X$, that can be considered in some sense
a ``coordinate operator''. This supposition is accepted as
undoubted in non-relativistic quantum mechanics. In contrast, in
relativistic quantum mechanics there is no natural notion of
coordinate.\footnote{This is true even for Dirac's electron.
Superposition of positive- and negative-frequency solutions of
Dirac's equation does not have any physical meaning, and therefore
there is no observable described by ``operator of multiplication
by variable $x$''.} It is possible, for example, to form from
generators of the Poincare group some vector operator, to which in
classical mechanics coordinates of the center of mass correspond.
But it should be understood that it will have some unusual
properties. For example, its coordinates will not commute with
each other.

Suppose that $X$ is a self-adjoint operator with continuous
spectrum, $X=X^*$. Let us denote $\Omega_{(a,b)}$ its spectral
projector\footnote{Roughly speaking, spectral projector is an
operator nullifying wave function in $X$-representation outside of
the given interval.} for an arbitrary real interval $(a,b)$.

Suppose also that $X$, being a coordinate, behaves in usual way
under action of translations, namely, for any $a$, $b$ and
 $\delta x$ we have equality:
\[
 e^{+\,i\,\delta x\,P/\hbar} \,
  \Omega_{(a+\delta x,b+\delta x)} \,
    e^{-\,i\,\delta x\,P/\hbar} \,=\,
 \Omega_{(a,b)} \ ,
\]
i.~e. $P$ is a ``generator of spectral shifts'' for $X$.

Suppose now that the system is in state $\rangle$,
$\langle\,|\,\rangle=1$.

The quantity $\langle\,\Omega_{(a,b)}\,\rangle$, obviously,
defines the probability to find the system inside the interval
$(a,b)$. Let us define such $l$ and $r$, that
\[
 \textstyle
 \langle\,\Omega_{(-\infty,l)}\,\rangle
 \,=\,
 \langle\,\Omega_{(r,+\infty)}\,\rangle
 \,=\,
 \frac{1-\sin1}2
 \,\approx\, 0,07926\dots
\]
It is easy to see, that $l$ and $r$ exist\footnote{But, generally
speaking, they are not unique. In order to eliminate this
non-uniqueness, it is convenient to choose $l$ maximum of the
possible, and $r$~--- minimum. Then the distance $r-l$ will be
minimum.}. So, the quantity $\delta_\rangle X = r-l$ can be
naturally called ``uncertainty'' of the coordinate $X$.

\mytheorem{}(uncertainty principle)
 {\it The following inequality takes place:
\begin{equation}
 \delta_\rangle X \Delta_\rangle P \geqslant \hbar \ .
 \label{Uncertainty}
\end{equation}}

In order to prove this theorem let us use (two times) the triangle
inequality for the quantum angle:
\[
 \angle\,\bigl(\
  \rangle
  \ , \
  e^{-\,i\,\delta_\rangle X\,P/\hbar}\,\rangle
 \ \bigr)
 \ \geqslant\
 \angle\,\bigl(\
  \Omega_{(-\infty,r)}\,\rangle
  \ , \
  \Omega_{(r,+\infty)}\,e^{-\,i\,\delta_\rangle X\,P/\hbar}\,\rangle
 \ \bigr)
 \,-\,
 \qquad\qquad\qquad\qquad\qquad\qquad
\]
\[
 \qquad\qquad\qquad\qquad\qquad\qquad
 \,-\,
 \angle\,\bigl(\
  \rangle
  \ , \
  \Omega_{(-\infty,r)}\,\rangle
 \ \bigr)
 \,-\,
 \angle\,\bigl(\
  e^{-\,i\,\delta_\rangle X\,P/\hbar}\,\rangle
  \ , \
  \Omega_{(r,+\infty)}\,e^{-\,i\,\delta_\rangle X\,P/\hbar}\,\rangle
 \ \bigr)
 \,=\,
\]
\[
 \textstyle
 \,=\,
 \frac\pi2
 \,-\,
 \arcsin\sqrt{\frac{1-\sin1}2}
 \,-\,
 \arcsin\sqrt{\frac{1-\sin1}2}
 \,=\,
 \frac\pi2
 \,-\,
 \left(\frac\pi4-\frac12\right)
 \,-\,
 \left(\frac\pi4-\frac12\right)
 \,=\,
 1 \ .
\]
But this means that under action of
 $e^{-\,i\,\delta_\rangle X\,P/\hbar}$
the vector $\rangle$ changes substantially.

Applying  the {\it certainty} principle, we directly get
(\ref{Uncertainty}). $\blacksquare$

So, the certainty principle allows to get the alternate inequality
(\ref{Uncertainty}) describing the uncertainty principle. This
inequality is not equivalent to the Kennard inequality
({\ref{KennardRel}).

From formal mathematical point of view, none of the inequalities
({\ref{KennardRel}) and (\ref{Uncertainty}) is stronger than the
other.

But it is not difficult to see that the fraction of the quantities
$\Delta_\rangle X$ and $\delta_\rangle X$ always satisfies the
inequality:
\[
 \frac{2\,\Delta_\rangle X}{\delta_\rangle X} \,\geqslant\,
 \sqrt{1-\sin1} \,\approx\, 0,398\dots
\]
Therefore, if we weaken the Kennard inequality ({\ref{KennardRel})
by the additional factor $0,398$ (which is usually not very
important for qualitative estimations), then it becomes weaker
than (\ref{Uncertainty}).

On the other hand, the fraction
 $\Delta_\rangle X/\delta_\rangle X$ can be arbitrarily great,
or even infinitely great, if the wave packet is badly localized.
In such a situation the Kennard inequality ({\ref{KennardRel}), in
contrast to the inequality (\ref{Uncertainty}), does not give any
estimation for the uncertainty of momentum at all.

So, the inequality (\ref{Uncertainty}) is more informative than
({\ref{KennardRel}) for qualitative estimations.


\paragraph{Space rotations.}
The generators of the three-parameter group of space rotations are
components of the vector operator of angular momentum $J_1$, $J_2$
and $J_3$. Consequently, the certainty relation is written as
\[
 \Delta_\rangle (\,
   \delta\varphi_1\,J_1\, +\,
   \delta\varphi_2\,J_2\, +\,
   \delta\varphi_3\,J_3\,
 ) \geqslant \hbar  \ .
\]
Logarithmic coordinates have here simple geometric interpretation:
rotation, described by coordinates $\delta\varphi_1$,
$\delta\varphi_2$ and $\delta\varphi_3$, can be performed as
rotation around the vector with these coordinates by angle equal
to the length of this vector.

If we suppose now that $\delta\varphi_2=\delta\varphi_3=0$ and
omit the index, then we have:
\[
 |\delta\varphi|\, \Delta_\rangle J \geqslant \hbar \ .
\]

Let us try to get an uncertainty relation by the same scheme as in
the case of the group of space translations.

First of all, it is necessary to note, that {\it uncertainty}
relation in the form ``product of uncertainty of angle and
uncertainty of angular momentum is greater or equal to Planck's
constant'' can not be suggested, because in a state, when
uncertainty of angular momentum is equal to zero (an eigenstate),
``uncertainty of angle'' should not be greater than $2\pi$.

Furthermore, here we meet the difficulty that it is impossible to
introduce a ``good'' self-adjoint operator of angle. The number of
operators turns out to be infinite (if at least one is defined),
but as soon as we fix the choice of one of them by some condition
(for example, that its spectrum is inside the interval from $0$ to
$2\pi$), it turns out to have quite complicated behaviour under
rotations, and analysis of formulas becomes difficult.

So, we will not choose here one from the infinite set of operators
of angle, but will imply that we talk about the whole infinite
set. For this infinite set we will use symbolic notation $\Phi$.

Such an approach does not meet difficulties, because when deriving
uncertainty relation we in fact operate with spectral measure (the
set of spectral projectors). And this spectral measure for all
operators of angle is the same in essence. Just argument of this
measure is not an interval of the real line, as in the case of
coordinate operator, but a segment of circle. And real coordinate
on the circle should be understand up to addition of $2\pi$. So,
the spectral measure satisfies relation:
\[
 \Omega_{(a+2\pi,b+2\pi)} = \Omega_{(a,b)} \ .
\]

Suppose also that the operator $J$ of angular momentum is a
``generator of spectral rotations'' for $\Phi$:
\[
 e^{+\,i\,\delta\varphi\,J/\hbar} \,
  \Omega_{(a+\delta\varphi,b+\delta\varphi)} \,
    e^{-\,i\,\delta\varphi\,J/\hbar} \,=\,
 \Omega_{(a,b)} \ .
\]

The quantity $\langle\,\Omega_{(a,b)}\,\rangle$, obviously,
defines the probability to find the system inside the angle
segment $(a,b)$. Suppose now that there is such a small segment
$(l,r)$, where mainly the probability to find the system is
concentrated. Namely, suppose that the probability to find the
system in the alternate segment is:
\[
 \textstyle
 \langle\,\Omega_{(r,l+2\pi)}\,\rangle
 \,=\,
 \frac{1-\sin1}2
 \,\approx\, 0,07926\dots
\]
If $r$ and $l$ were chosen so that the quantity
 $\delta_\rangle \Phi = r-l$ is minimum, then
$\delta_\rangle \Phi$ can be naturally called ``uncertainty'' of
the angle $\Phi$.

\mytheorem{} (uncertainty principle for angle and angular
momentum)
 {\it The following inequality takes place:
\begin{equation}
 \delta_\rangle\Phi \ \geqslant\
 \min\ (\ \hbar\,/\,\Delta_\rangle J\ ;\ \pi\ ) \ .
 \label{UncertRot}
\end{equation}}

Consider the case, when $\delta_\rangle\Phi\,\leqslant\,\pi$. In
this case the rotated angle segment $(l,r)$ will not overlap with
the non-rotated one, and we can apply the same estimation
technique for quantum angle as in the case of translations.

So, let us use (two times) the triangle inequality for the quantum
angle:
\[
 \angle\,\bigl(\
  \rangle
  \ , \
  e^{-\,i\,\delta_\rangle\Phi\,J/\hbar}\,\rangle
 \ \bigr)
 \ \geqslant\
 \angle\,\bigl(\
  \Omega_{(l,r)}\,\rangle
  \ , \
  \Omega_{(r,r+\delta_\rangle\Phi)}\,e^{-\,i\,\delta_\rangle\Phi\,J/\hbar}\,\rangle
 \ \bigr)
 \,-\,
 \qquad\qquad\qquad\qquad\qquad\qquad
\]
\[
 \qquad\qquad\qquad\qquad\qquad\qquad
 \,-\,
 \angle\,\bigl(\
  \rangle
  \ , \
  \Omega_{(l,r)}\,\rangle
 \ \bigr)
 \,-\,
 \angle\,\bigl(\
  e^{-\,i\,\delta_\rangle\Phi\,J/\hbar}\,\rangle
  \ , \
  \Omega_{(r,r+\delta_\rangle\Phi)}\,e^{-\,i\,\delta_\rangle\Phi\,J/\hbar}\,\rangle
 \ \bigr)
 \,=\,
\]
\[
 \textstyle
 \,=\,
 \frac\pi2
 \,-\,
 \arcsin\sqrt{\frac{1-\sin1}2}
 \,-\,
 \arcsin\sqrt{\frac{1-\sin1}2}
 \,=\,
 \frac\pi2
 \,-\,
 \left(\frac\pi4-\frac12\right)
 \,-\,
 \left(\frac\pi4-\frac12\right)
 \,=\,
 1 \ .
\]
But this means that under action of
 $e^{-\,i\,\delta_\rangle\Phi\,J/\hbar}$
the vector $\rangle$ is changed substantially.

Applying the {\it certainty} principle, we directly get
(\ref{UncertRot}). $\blacksquare$

We see that the certainty principle allows to easily get also the
simple {\it uncertainty} relation for angle and angular momentum.

It should be noted that attempt \cite{Judge1964} to get an
uncertainty relation for angle and angular momentum by analogy
with the Kennard inequality ({\ref{KennardRel}) leads to quite
complicated formulas. Namely, if we fix the self-adjoint operator
of angle $\Phi$ by the condition, that its spectrum is the
interval $[-\pi,\pi]$, and define the ``uncertainty'' of angle by
the formula:
\[
 \Delta'_\rangle\Phi =
  \min_{\delta\varphi}
  \sqrt{\langle\,
   e^{+\,i\,\delta\varphi\,J/\hbar} \,
   \Phi^2 \,
   e^{-\,i\,\delta\varphi\,J/\hbar}\,\rangle} \ ,
\]
then we have the uncertainty relation:
\begin{equation}
 \Delta'_\rangle\Phi \, \Delta_\rangle J \,\geqslant\,
 \frac\hbar2
 \,
 \left[\,
  1\, -\, \frac3{\pi^2}\, (\Delta'_\rangle\Phi)^2
 \,\right]
 \ .
 \label{UncJudge}
\end{equation}
For investigation of semiclassical limit the inequality
(\ref{UncertRot}) turns out to be more convenient than
(\ref{UncJudge}), because allows to use weaker conditions for
localization of wave packet.


\paragraph{Time shifts.}
Poincare group includes as a subgroup one-parameter group of
shifts in time. It is natural to expect that the certainty
principle has corresponding mathematical expression in this case.
And this is really so.

But our discussion here is complicated by the circumstance that
all educational literature on both classical and quantum mechanics
is oriented to resolving of problems of dynamics only. And the
notion of dynamical state is necessarily attributed to some moment
of time. Such approach hides relativistic invariance. So, the
notion of time shifts becomes senseless.

Completely self-consistent, from this point of view, are invariant
Hamiltonian formalism (in the case of the classical mechanics) and
relativistic canonical quantization (RCQ) (in the case of
relativistic quantum field theory). In these approaches the theory
includes not only time shifts, but also Lorentz boosts. But, so
far as most physicists are not well-acquainted with such
approaches now, here we will discuss some intermediate approach
that is in essence based on Heisenberg representation. (Lorentz
boosts  we will not consider in this review at all.)

Usually, dynamical state $S_t$ in some moment of time $t$ is
understood as some set of values, that can be measured in the
moment $t$ by some set of apparatuses, and that define the whole
subsequent evolution of the system (i.~e. analogous sets of values
in subsequent moments of time). In classical mechanics dynamical
state is usually described either by set of coordinates and
velocities (Lagrangian formalism) or by set of coordinates and
momenta (Hamiltonian formalism). In other words, dynamical state
in a given moment of time is defined by a point on some manifold,
which is called phase space. In quantum mechanics dynamical state
is described by vector in Hilbert space.

When we have the question about description of the group of space
motions on the space of dynamical states in some moment of time,
we usually think in the following way. Suppose a set of physical
apparatuses $K$, which measures the full set of dynamical
variables of the system under consideration, under action of some
motion $G$ transferred to $K'$, $G:K\to K'$. And suppose for some
dynamical state of the system $S$ we have dynamical state $S'$,
such that the set of apparatuses $K'$, which measures parameters
of the state $S'$, gives the same results as $K$, which measures
parameters of the state $S$. Then we say that under action of the
transformation $G$ state $S$ transfers to $S'$. Symbolically we
can write it as:
\[
 G:K\to K' \ ,\ K(S)=K'(S')
 \qquad\Longrightarrow\qquad
 G:S\to S' \ .
\]
Exactly in this way the action of space translations and rotations
is introduced, described above.

It seems, it would be natural to apply this definition to the
group of time shifts:
\[
 G:K_{t_1}\to K_{t_2} \ ,\ K_{t_1}(S_{t_1})=K_{t_2}(S_{t_2})
 \qquad\Longrightarrow\qquad
 G:S_{t_1}\to S_{t_2} \ .
\]
But this definition turns out to be not very useful, because it
turns out to be purely formal and with no connection with
dynamics.

Let us improve a little the definition of dynamical state given
above. Let us suppose that if some dynamical state $S_{t_1}$ is
transferred by evolution to $S_{t_2}$, then $S_{t_1}$ and
$S_{t_2}$ describe the same dynamical state. In other words,
dynamical state is just some abstract set of values (not
necessarily connected with measurements in a concrete moment of
time) defining evolution of the system. When we talk about quantum
systems, we can think that dynamical state is just described by
Heisenberg vector of state.

If the system is closed, its Hamiltonian does not depend on time,
and it commutes with space motions. Therefore the theory of space
motions, described in the previous sections, can be applied in
this case. And what is more, the group of time shifts is joined
here, which is described by the family of operators:
\[
 U(\delta t) = e^{-\,i\,\delta t\,(-H)/\hbar} \ .
\]
It should be noticed that generator of this group is Hamiltonian
with sign ``minus''. An operator, describing action of the group
under consideration, is reciprocal of the operator, describing
evolution in Schr\"odinger equation. We can say that here the
group acts actively, but in the Schr\"odinger representation it
acts passively.

The certainty relation for the group of time shifts takes the
form:
\begin{equation}
 |\delta t|\, \Delta_\rangle H \geqslant \hbar \ .
 \label{CertTH}
\end{equation}
If we look at this relation from the point of view of the
Schr\"odinger representation (but doing so we lose connection with
Poincare group), then this relation turns out to be exactly the
Mandelshtam-Tamm relation \cite{MandTamm1945} for the speed of
quantum evolution.

So, the Mandelshtam-Tamm relation for the case of closed systems
is a consequence of the certainty principle.

Let us consider now non-closed systems. The Mandelshtam-Tamm
relation for the speed of quantum evolution (derived in the
Schr\"odinger representation) is naturally generalized for this
case. Let us denote Hamiltonian for this case as
$H^{\mathrm{full}}$. So far as the quantity
 $\Delta_{|t\rangle} H^{\mathrm{full}}$, generally speaking,
depends on time, it should be just averaged:
\begin{equation}
 |\tau|\, \overline{\Delta_{|t\rangle} H^{\mathrm{full}}}
 \geqslant \hbar \ .
 \label{MTFull}
\end{equation}
Here $\tau$ is the time interval of evolution (it does not have
meaning of logarithmic coordinate on any group).

The certainty principle in this situation, formally, turns out to
be not applicable. But, in practical situations, when we consider
non-closed quantum systems, their Hamiltonian can be usually
represented as sum of two terms, the Hamiltonian of ``free''
system and Hamiltonian of interaction ``with external classical
field'':
\[
 H^{\mathrm{full}} = H + H^{\mathrm{int}} \ .
\]
And usually the Hamiltonian of interaction $H^{\mathrm{int}}$ can
be ``turned off''.

In this case it is natural to connect the notion of dynamical
state of the system with its evolution when Hamiltonian of
interaction is turned off (though, the state can be considered in
a concrete moment of time, without turning off the Hamiltonian of
interaction). Correspondingly, the certainty relation
(\ref{CertTH}) remains the same. Its meaning becomes somewhat more
abstract, with no direct connection with the Mandelshtam-Tamm
relation (\ref{MTFull}).

In the end, let us note that Mandelshtam and Tamm derived their
relation by direct estimation of the scalar product of the initial
and final vectors of state. They did not use the theory of
Fubini-Study metric. Such a derivation turns out to be
substantially shorter. But geometrical nature of the result turns
out to be quite hidden. This additionally complicates
interpretation (in addition to the more serious disadvantage:
interpretation based on the Schr\"odinger equation).

Derivation of the Mandelshtam-Tamm relation (in the frame of the
Schr\"odinger equation), based on Fubini-Study metric, was given
later by Anandan and Aharonov \cite{AnanAhar1990}. They did not
give the reference to Mandelshtam and Tamm. So, in the western
literature this relation is often erroneously called the
Anandan-Aharonov relation.

\section*{Acknowledgements}

In closing I want to thank A.~Kleyn, S.~Lorenz, T.~A.~Bolokhov,
A.~V.~Ossipov, E.~V.~Aksyonova, A.~Yu.~Tos\-che\-vi\-ko\-va,
M.~Gatti, and A.~K.~Pati for helpful discussions, references and
help in procuring literature.


\section*{}

\tcptab


\end{document}



\title{Принцип определённости (обзор)}
\author{Д.~А.~Арбатский\footnote{ http://daarb.narod.ru/ , http://wave.front.ru/ }}
\date{Август 2006 г.}
\maketitle

\begin{abstract}
Принцип определённости (2005) позволил осмыслить с более
фундаментальных позиций как принцип неопределённости Гейзенберга
(1927), так и соотношение Мандельштама-Тамма (1945). В настоящем
обзоре даётся подробное объяснение и обсуждение принципа
определённости, ориентированное на всех физиков, как теоретиков,
так и экспериментаторов.
\end{abstract}

\section*{Исторические комментарии}

\paragraph{Принцип неопределённости.}
Читателю, интересующемуся историей принципа {\it
неопределённости}, рекомендуется познакомиться, например, с
обзором \cite{HilgUff2001}. Здесь же будут даны только некоторые
замечания, важные с точки зрения дальнейшего изложения.

Принцип неопределённости был предложен Гейзенбергом в 1927~г.
\cite{Heis1927}. Сам Гейзенберг его сформулировал так:
\begin{itemize}
\item
Чем более точно определена координата, тем менее точно в то же
время известен импульс, и наоборот.
\end{itemize}
При этом для неточностей координаты и импульса было указано
качественное соотношение вида:
\begin{equation}
 \Delta_\rangle X\,\Delta_\rangle P \,\sim\, \hbar \ .
 \label{HeisRel1}
\end{equation}

Гейзенберг, рассматривая конкретные примеры, дал именно
качественную формулировку. При этом самому понятию \rql
неточности\rqr{} он не дал точного определения.

Вскоре Кеннард \cite{Kennard1927} дал точную математическую
формулировку для случая координаты и импульса. Предполагая
справедливость \rql коммутационного соотношения\rqr{}
$[X,P]=i\hbar$, он показал, что для произвольного квантового
состояния $\rangle$ имеет место соотношение:
\begin{equation}
 \Delta_\rangle X\,\Delta_\rangle P \,\geqslant\,
 \frac\hbar2 \ .
 \label{KennardRel}
\end{equation}
При этом
\[
 \Delta_\rangle X =
   \left\langle\bigl(\,
    X-\langle X\rangle
   \,\bigr)^2\right\rangle^{1/2}
 \ , \qquad
 \Delta_\rangle P =
   \left\langle\bigl(\,
    P-\langle P\rangle
   \,\bigr)^2\right\rangle^{1/2}
 \ .
\]
То есть, в соответствии с вероятностной интерпретацией квантовой
механики, под неточностями координаты и импульса предлагалось
понимать среднеквадратичные отклонения этих наблюдаемых.

Ввиду исключительной простоты доказательства, соотношение
(\ref{KennardRel}) стало общепризнанным математическим выражением
принципа неопределённости и попало во все учебники квантовой
механики. При этом всевозможная критика (в том числе в плане
соответствия практическому эксперименту) \cite{HilgUff2001}
оказалась более или менее проигнорированной. Однако, скоро мы
увидим, что принцип определённости позволяет получать
альтернативные неравенства, описывающие принцип неопределённости,
которые оказываются более содержательными с практической точки
зрения.

Далее математическое доказательство различным образом обобщалось.
При этом оно распространялось на другие пары некоммутирующих
наблюдаемых. Поэтому, чтобы избежать неясности в самом термине
{\it принцип неопределённости}, приведём здесь несколько иную,
более современную и общую, формулировку этого
принципа\footnote{Такая формулировка не только позволяет яснее
понять, что же именно является \rql неопределённым\rqr{}, но и
позволяет увидеть ясно контраст с принципом {\it
определённости}.}:
\begin{itemize}
\item
Если попытаться описывать динамическое состояние {\it квантовой}
частицы методами {\it классической} механики, то точность такого
описания принципиально ограничена. {\it Классическое} состояние
частицы оказывается {\it плохо} определённым. Эта неопределённость
может быть математически выражена различными неравенствами,
описывающими разброс значений наблюдаемых, имеющих
квазиклассический предел.
\end{itemize}

Разумеется, Гейзенберг не мог сформулировать принцип
неопределённости в такой форме. Это бы противоречило логике
исторического момента. Ведь в 1927 году физика переходила от
классической механики к квантовой, и принцип неопределённости
воспринимался как путь из старой теории в новую.

Сегодня же, когда квантовая механика уже состоялась, принцип
неопределённости является всего лишь методом качественной оценки
точности квазиклассического приближения. И нет никаких оснований
считать этот принцип более фундаментальным, чем сам математический
аппарат квантовой механики.


\paragraph{Соотношение Мандельштама-Тамма.}
Поскольку для координаты и импульса существует соотношение
неопределённостей вида (\ref{HeisRel1}), общие соображения,
связанные с теорией относительности,\footnote{Скоро мы, однако,
увидим, что эти соображения {\it неверны}, ибо само существование
принципа неопределённости связано со специфическими особенностями
{\it нерелятивистского} приближения.} указывают, что, по-видимому,
должно существовать и соотношение неопределённостей для времени и
энергии вида:
\begin{equation}
 \Delta_\rangle T\,\Delta_\rangle H \,\sim\, \hbar \ .
 \label{TH1}
\end{equation}
Сам Гейзенберг указал на такое соотношение. Однако, уже в то время
было ясно \cite{Pauli1933}, что с данным соотношением дело обстоит
сложнее, так как не существует квантовомеханической наблюдаемой
\rql время\rqr{}, и не ясно, что же следует понимать под \rql
неопределённостью\rqr{} времени.

Тем не менее, физикам всё же хотелось верить, что принцип
неопределённости является {\it фундаментальным} физическим
принципом, а потому попытки обосновать соотношение (\ref{TH1}) не
прекращались до самого последнего времени. При этом они в основном
сводились к попыткам показать, что понятие \rql неопределённость
времени\rqr{} является осмысленным. И это осмысление обычно
проводилось путём анализа процесса измерения, то есть к анализу
взаимодействия квантовой частицы с прибором типа микроскопа
Гейзенберга. (подробный обзор с критическим анализом см. в
\cite{Voron1981})

Нельзя сказать, что эти попытки были безуспешными. Тем не менее,
думается, такой подход не соответствует общей методологии
квантовой механики. Дело в том, что аналитический аппарат
квантовой механики ориентирован на изучение квантовых систем как
самостоятельных физических сущностей. Этот аналитический аппарат
дистанцируется от подробностей взаимодействия приборов с изучаемой
системой.

В 1945 г. Мандельштам и Тамм \cite{MandTamm1945} дали строгую
математическую формулировку соотношения вида\footnote{Здесь оно
записано несколько иначе, чтобы яснее была видна связь с
дальнейшим.}
\begin{equation}
 |\delta t|\, \Delta_\rangle H \geqslant \hbar \ .
 \label{MandTammRel}
\end{equation}
Мандельштам и Тамм назвали это соотношение \rql соотношением
неопределённости энергия-время в нерелятивистской квантовой
механике\rqr{}, поскольку, очевидно, полагали, что нашли
математическое выражение принципа неопределённости для энергии и
времени. Соответственно, дальнейшие попытки обосновать соотношение
{\it неопределённости} энергия-время в основном являлись попытками
показать, что величина $\delta t$ в соотношении
Ман\-дель\-шта\-ма-Тамма может в том или ином смысле пониматься
как \rql неопределённость\rqr{} времени.

Однако, я всё же полагаю, что называть соотношение
Мандельштама-Тамма соотношением {\it неопределённостей} нет
достаточно убедительных оснований. Как было сказано выше, принцип
неопределённости не более фундаментален, чем сам математический
аппарат квантовой механики. Более того, анализ релятивистских
квантовых систем показывает, что он {\it менее} фундаментален.
Поэтому нет оснований считать, что фундаментальное соотношение
неопределённостей вида (\ref{TH1}) должно существовать.

Скоро мы увидим, что соотношение Мандельштама-Тамма в случае
замкнутых систем является проявлением иного физического принципа,
принципа {\it определённости}, который более фундаментален, чем
принцип неопределённости Гейзенберга. При этом величина
 $\delta t$, фигурирующая в неравенстве (\ref{MandTammRel}),
при переходе от представления Шрёдингера к представлению
Гейзенберга (или к РКК-представлению) оказывается логарифмической
координатой на группе Пуанкаре. Для некоторых целей её можно
интерпретировать как \rql неопределённость времени\rqr{}, но в
этом, вообще говоря, нет нужды.

Что касается незамкнутых систем, у которых гамильтониан
произвольно зависит от времени, соотношение Мандельштама-Тамма
следует считать соотношением для скорости квантовой эволюции. При
этом оно является вполне самостоятельным соотношением,
математически {\it не}-эквивалентным соотношению определённости
для времени и энергии.


\paragraph{Принцип определённости.}
Когда встаёт вопрос о применении соображений, связанных с теорией
относительности, в квантовой механике, стоит задуматься о том, что
нам вообще известно про релятивистскую инвариантность в квантовой
теории.

Группой инвариантности пространства
Минковского\footnote{Четырёхмерного пространства-времени
специальной теории относительности.} является группа Пуанкаре.
Именно с инвариантностью по отношению к этой группе законов
природы вообще, и квантовой механики в частности, связывается
понятие релятивистской инвариантности.

Все физически содержательные квантовые системы, на которые
распространяется действие группы Пуанкаре, известные до сих пор,
являются квантовыми полями (либо их подсистемами). При этом до
2002 года математическое конструирование квантовых полей
производилось по таким рецептам\footnote{Именно \rql
рецептам\rqr{}, поскольку формальный математический процесс вообще
отсутствовал.}, которые делали действие группы Пуанкаре на
пространстве состояний квантованного поля скрытым. Фактически,
физики довольствовались утверждением, что группа Пуанкаре
действует на пространстве состояний {\it неявно}, по некоторым
довольно сложным формулам.

С появлением релятивистского канонического квантования (РКК)
\cite{Arb2002} (для начального ознакомления рекомендуется обзор
\cite{Arb2005wircq}) стало, однако, понятно, каким образом
действие группы Пуанкаре переносится с пространства Минковского на
пространство состояний квантованного поля.\footnote{Отличительной
особенностью метода РКК является именно построение квантованных
полей в рамках четырёхмерной геометрии пространства Минковского,
без искусственного разделения пространства и времени.} Стало также
очевидным следующее:
\begin{itemize}
\item
Понятие координаты, как квантовомеханической наблюдаемой, для
релятивистских квантовых систем вообще не является естественным,
не говоря уже о фундаментальности. (Предпринимавшиеся в прошлом
попытки введения координат для некоторых систем, например
\cite{NewtWig1949}, представляются довольно искусственными.)
\item
По этой причине принцип неопределённости вообще {\it не может}
быть фундаментальным, в плане распространения его на
релятивистскую квантовую теорию.
\end{itemize}

Опираясь на методологию РКК, я сформулировал \cite{Arb2005tcp}
принцип {\it определённости}:
\begin{itemize}
\item
Если описывать динамическое состояние квантовой частицы (системы)
методами {\it квантовой} механики, то {\it квантовое} состояние
частицы (системы) оказывается {\it хорошо} определённым. Эта
определённость квантового динамического состояния означает, что
\rql малые\rqr{} пространственно-временные преобразования не могут
существенно менять квантовое состояние. При этом в случае группы
Пуанкаре, для преобразований, способных существенно менять
квантовое состояние, справедлива оценка\footnote{Здесь и далее по
повторяющимся греческим тензорным индексам предполагается
релятивистское суммирование:
\[
 a_\mu\, b_\mu = a_0\, b_0 - a_1\, b_1 - a_2\, b_2 - a_3\, b_3 \ .
\].}:
\begin{equation}
 \textstyle
 \Delta_\rangle\,
   (\,-\,\delta x_\mu\, P_\mu\, +
      \frac12\,
      \delta\omega_{\mu\nu}\, J_{\mu\nu} \,) \geqslant \hbar \ ,
 \label{CertPoin}
\end{equation}
где $P_\mu$~--- векторный оператор энергии-импульса,
$J_{\mu\nu}$~--- тензорный оператор четырёхмерного момента
импульса, $\delta x_\mu$ и $\delta\omega_{\mu\nu}$~--- стандартные
логарифмические координаты группы Пуанкаре.
\end{itemize}

При этом, поскольку принцип определённости является универсальным,
он прекрасно работает и в обычной квантовой механике, в том числе
в {\it нерелятивистской}. Поэтому кажется удивительным, что этот
принцип не был сформулирован раньше. Этому, однако, есть два
объяснения.

Во-первых, этот принцип формулируется естественно с использованием
базовых понятий РКК. В частности, при формулировке принципа
определённости понятие {\it динамического состояния} наиболее
естественно понимать в интерпретации РКК. Эта трудность,
по-видимому, являлась главной.

Во-вторых, для того, чтобы прийти к принципу определённости из
обычной квантовой механики, требуется опираться на теорию
квантового угла (метрики Фубини-Штуди). А существование этой
метрики, хотя и несложно доказывается, отнюдь не является
очевидным. Эта метрика была открыта только в 1905 году Фубини
\cite{Fubini1904} и Штуди \cite{Study1905}, и большинство физиков
о ней не знают.

\section*{Математический формализм}

\paragraph{Квантовый угол (метрика Фубини-Штуди).}
Рассмотрим единичную сферу $\|x\|=1$ в обычном трёхмерном
евклидовом пространстве $\mathbb R^3$. Расстояние, измеренное по
поверхности сферы, между любыми двумя точками равно углу между
лучами, исходящими из начала координат и проходящими через данные
точки.

Таким образом, связь углов между лучами с расстояниями на сфере
делает очевидным важное свойство угла: он удовлетворяет \rql
неравенству треугольника\rqr{}:
\begin{equation}
 \angle(a,c) \leqslant \angle(a,b) + \angle(b,c) \ .
 \label{TrianIneq}
\end{equation}

Угол в этом примере, как известно, выражается через скалярное
произведение $\langle\cdot|\cdot\rangle$ в пространстве $\mathbb
R^3$ по формуле:
\[
 \angle(a,b)\, = \,\arccos\, \langle a|b\rangle \ .
\]
Здесь в правой части формулы векторы, концы которых лежат в точках
пересечения лучей $a$ и $b$ с единичной сферой, обозначены теми же
буквами, что и сами лучи. Возможные значения угла в этом примере
лежат в диапазоне от $0$ до $\pi$.

Попробуем теперь рассмотреть углы не между лучами, а между
прямыми, проходящими через начало координат. При этом возможные
значения угла, очевидно, будут лежать в более узком диапазоне от
$0$ до $\pi/2$. Формула для угла в этом случае немного
видоизменяется:
\begin{equation}
 \angle(a,b)\, = \,\arccos\,\bigl| \langle a|b\rangle \bigr| \ .
 \label{QAngle}
\end{equation}

Будет ли угол в этом случае удовлетворять неравенству треугольника
(\ref{TrianIneq})? Ответ на этот вопрос не кажется непосредственно
очевидным, поскольку каждой прямой отвечает уже пара точек на
единичной сфере. Однако, элементарное логическое рассуждение
позволяет свести этот пример к предыдущему. Ответ оказывается
положительным.

Попробуем теперь посмотреть, каким образом обстоят дела в
трёхмерном комплексном гильбертовом пространстве $\mathbb C^3$.
Вместо вещественных прямых в этом случае мы, очевидно, имеем
комплексные прямые, то есть одномерные комплексные
подпространства. Каждое такое подпространство пересекается с
единичной сферой $\|x\|=1$ уже даже не по двум, а по бесконечному
числу точек. Эти точки отличаются друг от друга комплексным
множителем, равным по модулю единице.

Угол при этом определяется той же самой формулой (\ref{QAngle}),
но скалярное произведение здесь уже соответствует пространству
$\mathbb C^3$. Возможные значения угла, очевидно, также будут
лежать в диапазоне от $0$ до $\pi/2$.

Будет ли угол и в этом случае удовлетворять неравенству
треугольника (\ref{TrianIneq})? Ответ и на этот раз оказывается
положительным. Этот факт был установлен Фубини \cite{Fubini1904} и
Штуди \cite{Study1905} (несложное доказательство также имеется в
\cite{Arb2005tcp}), поэтому угол между комплексными прямыми в
гильбертовом пространстве называется {\it метрикой Фубини-Штуди}.

Выполнение неравенства треугольника, конечно, не зависит от
размерности гильбертова пространства, поскольку при проверке
неравенства треугольника всегда можно перейти к соответствующему
трёхмерному подпространству.

Далее, поскольку в квантовой механике пространство состояний
квантовой системы $\mathcal H$ является комплексным гильбертовым
пространством, метрика Фубини-Штуди может естественным образом
рассматриваться как мера различности квантовых состояний. В этом
контексте мы будем называть данную метрику {\it квантовым углом},
поскольку это позволяет образовывать естественные словосочетания
типа \rql квантовая угловая скорость\rqr{}.


\paragraph{Квантовая угловая скорость.}
Пусть теперь нормированный вектор $r$ зависит от вещественного
параметра $t$: $t\in\mathbb R$, $r(t)\in\mathcal H$,
$\|\,r(t)\,\|=1$.

Определим {\it квантовую скорость} $v(t)$ равенством:
\[
 v(t) = \dot r(t) =
    \lim_{\delta t\to0}
    \frac{r(t+\delta t)-r(t)}{\delta t} \ .
\]

Определим также {\it квантовую угловую скорость} $\omega(t)$
равенством:
\[
 \omega(t) = \lim_{\delta t\to0}
     \frac{\angle\,\big(\,r(t+\delta t)\,,\,r(t)\,\big)}{|\,\delta t\,|} \ .
\]

Чтобы выразить $\omega(t)$ через $v(t)$, разложим $v(t)$ на две
ортогональные составляющие:
\[
 v_\parallel(t) = r(t)\,\langle r(t)|v(t)\rangle \ , \qquad
 v_\perp(t) = v(t) - v_\parallel(t) \ .
\]

Данное разложение формально выглядит так же, как и соответствующее
разложение в вещественном случае, хорошо известное в кинематике
материальной точки. Однако, на эту аналогию нельзя полагаться
слишком сильно. Дело в том, что в вещественном случае наличие
параллельной компоненты скорости непременно связано с приращением
нормы вектора. В рассматриваемом же нами комплексном случае вектор
предполагается всегда сохраняющим единичную норму. Наличие же
параллельной компоненты скорости оказывается связанным с
возможностью домножения вектора на комплексный фазовый множитель,
равный по модулю единице. Тем не менее, верна следующая

\mytheorem{}
 {\it Квантовая угловая скорость равна норме ортогональной
составляющей квантовой скорости:
\begin{equation}
 \omega(t)=\|\,v_\perp(t)\,\| \ .
 \label{QAngSpeedQVel}
\end{equation}}

Прежде чем доказывать эту теорему, заметим, что в вещественном
случае равенство (\ref{QAngSpeedQVel}) является {\it определением}
угловой скорости $\omega$. При этом угол в вещественном случае
{\it определяется} как интеграл от угловой скорости вдоль
экстремальной дуги. После чего уже {\it определяется}
тригонометрическая функция $\arccos$ для {\it конечных} углов.

Непосредственное перенесение такой логики на случай комплексного
гильбертова пространства затруднительно, поскольку экстремальных
дуг всегда много, и ни одна из них не является столь простой,
чтобы появление вещественной функции $\arccos$ было геометрически
очевидно.

Поэтому здесь нам приходится исследовать непосредственно
бесконечно малые углы, т.~е. работать вблизи особой точки функции
$\arccos$. Чтобы этого избежать, мы используем равенство Парсеваля
для замены $\arccos$ на $\arcsin$:
\[
 \angle\,\big(\,r(t+\delta t)\,,\,r(t)\,\big) \,=\,
 \arccos\,\big|\,\langle\,r(t+\delta t)\,|\,r(t)\,\rangle\,\big| \,=\,
\]
\[
 \,=\,
 \arcsin\,
   \big\|\,
   r(t+\delta t) -
   r(t) \,\langle\, r(t) \,|\, r(t+\delta t) \,\rangle
   \,\big\| \,=\,
\]
\[
 \,=\,
   \arcsin\,
    \big\|\,
    r(t+\delta t) \,-\, r(t) \,-\,
    r(t) \,
     \langle\,
        r(t) \,|\, r(t+\delta t)-r(t)
     \,\rangle
    \,\big\|
 \,=\,
\]
\[
 \,=\,
   \arcsin\,
    \big\|\,
    v(t)\,\delta t \,+\, o(\delta t) \,-\,
    r(t) \,
     \langle\,
        r(t) \,|\, v(t)\,\delta t + o(\delta t)
     \,\rangle
    \,\big\|
 \,=\,
\]
\[
 \,=\,
   \arcsin\,
    \Big\|\,
     \big(v(t)\,-\,
     r(t) \,
     \langle\,
        r(t) \,|\, v(t)
     \,\rangle\big)
     \,\delta t \,+\, o(\delta t)
    \,\Big\|
 \,=\,
\]
\[
 \,=\,
   \arcsin\,
    \big\|\,
    v_\perp(t)\,\delta t \,+\, o(\delta t)
    \,\big\|
 \,=\,
   \arcsin\,
    \big(\,
    \|v_\perp(t)\|\,|\delta t| \,+\, o(\delta t)
    \,\big)
 \,=\,
\]
\[
 \,=\,
 \|v_\perp(t)\|\,|\delta t| \,+\, o(\delta t) \ .
\]
Деля на $|\delta t|$, получаем равенство (\ref{QAngSpeedQVel}).
$\blacksquare$


\paragraph{Непрерывные унитарные группы.}
Допустим теперь, что в пространстве состояний квантовой системы
имеется самосопряжённый оператор $A=A^*$. Множество унитарных
операторов вида $U(\delta s)=e^{-\,i\,\delta s\,A/\hbar}$, где
$\delta s$ пробегает всё множество вещественных чисел, а $\hbar$
является фиксированным положительным вещественным числом (в
квантовой механике это число является постоянной Планка), образует
сильно непрерывную однопараметрическую унитарную группу:
 $U(\delta s_1)\,U(\delta s_2)\,=\, U(\delta s_1 + \delta s_2)$.
Оператор $A$ при этом называется {\it генератором} данной группы.

И пусть теперь вектор состояния изменяется под действием
рассматриваемой группы:
\[
 r(\delta s) \,=\, |\,\delta s\,\rangle \,=\,
 U(\delta s)\,\rangle \,=\,
 e^{-\,i\,\delta s\,A/\hbar} \,\rangle \ .
\]
Здесь $|\,\delta s\,\rangle\in\mathcal H$~--- другое обозначение
вектора состояния при параметре, равном $\delta s$;
$\rangle\in\mathcal H$~--- фиксированный кет-вектор состояния.

Будем полагать, что функция $r(\delta s)$ дифференцируема. Тогда
квантовая скорость выражается формулой:
\[
 \textstyle
 v(\delta s) \,=\,
 \frac1{i\hbar}\, A \, e^{-\,i\,\delta s\,A/\hbar} \,\rangle \,=\,
 \frac1{i\hbar}\, A \, |\,\delta s\,\rangle \ .
\]

Среднее значение оператора $A$ не зависит от параметра:
\[
 \overline A \,=\,
 \langle\,\delta s\,|\, A \, |\,\delta s\,\rangle \,=\,
 \langle\,e^{+\,i\,\delta s\,A/\hbar} \, A \,
    e^{-\,i\,\delta s\,A/\hbar} \,\rangle \,=\,
 \langle\, A \, e^{+\,i\,\delta s\,A/\hbar} \,
    e^{-\,i\,\delta s\,A/\hbar} \,\rangle \,=\,
 \langle\, A \, \rangle \ .
\]

Поэтому составляющие квантовой скорости могут быть просто записаны
как:
\[
 \textstyle
 v_\parallel(\delta s) \,=\,
 |\,\delta s\,\rangle \, \langle\,\delta s\,|\,
 \frac1{i\hbar}\, A \, |\,\delta s\,\rangle \,=\,
 \frac1{i\hbar}\,\overline A \, |\,\delta s\,\rangle
\]
\[
 \textstyle
 v_\perp(\delta s) \,=\,
 \frac1{i\hbar}\,\left(A-\overline A\right) \,
 |\,\delta s\,\rangle \ .
\]

Квантовая угловая скорость оказывается также независящей от
времени:
\[
 \textstyle
 \omega(\delta s)
 \,=\,
 \|\,v_\perp(\delta s)\,\|
 \,=\,
 \frac1\hbar\,
 \langle\,\delta s\,|\,
    \left(A-\overline A\right)^2
 \, |\,\delta s\,\rangle^{1/2}
 \,=\,
 \frac1\hbar\,
 \langle\,
    e^{+\,i\,\delta s\,A/\hbar} \,
    \left(A-\overline A\right)^2 \,
    e^{-\,i\,\delta s\,A/\hbar}
 \,\rangle^{1/2}
 \,=\,
\]
\[
 \textstyle
 \,=\,
 \frac1\hbar\,
 \langle\,
    \left(A-\overline A\right)^2 \,
    e^{+\,i\,\delta s\,A/\hbar} \,
    e^{-\,i\,\delta s\,A/\hbar}
 \,\rangle^{1/2}
 \,=\,
 \frac1\hbar\,
 \langle\,
    \left(A-\overline A\right)^2
 \,\rangle^{1/2}
 \,=\,
 \frac1\hbar\, \Delta_\rangle A  \ .
\]
Здесь $\Delta_\rangle A$ является сокращённым обозначением для
среднеквадратичного отклонения $A$ в состоянии $\rangle$.

Таким образом, доказана важная

\mytheorem{}
 {\it Среднеквадратичное отклонение генератора однопараметрической
унитарной группы равно квантовой угловой скорости, умноженной на
постоянную Планка $\hbar$. При этом обе эти величины при действии
группы остаются неизменными.}

Посмотрим теперь, насколько сильно могут разойтись начальный
$\rangle$ и конечный $|\,\delta s\,\rangle$ векторы. Неравенство
треугольника обычным образом обобщается на произвольные
криволинейные пути \cite{Arb2005tcp}. Поэтому угол между начальным
и конечным векторами не может быть больше, чем приращение
параметра, умноженное на квантовую угловую скорость:
\begin{equation}
 \textstyle
 \angle\,\big(\,|\,\delta s\,\rangle\,,\,\rangle\,\big) \,\leqslant\,
 |\delta s| \, \frac1\hbar\, \Delta_\rangle A  \ .
 \label{AngEst}
\end{equation}

Для физических приложений представляется удобным ввести некоторую
качественную границу, при которой два вектора состояния могут
считаться различающимися {\it существенно}. Поэтому мы будем
говорить, что два вектора различаются {\it существенно}, если угол
между ними больше либо равен $1$.

Переформулируем теперь неравенство (\ref{AngEst}) в форме
следующей теоремы.

\mytheorem{} (принцип определённости)
 {\it Чтобы под действием сильно непрерывной однопараметрической унитарной
группы $U(\delta s)=e^{-\,i\,\delta s\,A/\hbar}$ начальный вектор
состояния $\rangle$ изменился существенно, необходимо, чтобы
выполнялось неравенство:
\begin{equation}
 |\delta s|\, \Delta_\rangle A \geqslant \hbar \ .
 \label{Certainty1}
\end{equation}}

Эта теорема легко обобщается и на случай, когда группа является
многопараметрической. Действительно, если $\delta s$ и $A$ имеют
матричные индексы, по которым предполагается суммирование, то
неравенство (\ref{Certainty1}) для этого случая можно записать
немного иначе:
\[
 \Delta_\rangle (\,\delta s_j\, A_j\,) \geqslant \hbar \ .
\]
При этом величины $\delta s_j$ являются так называемыми
логарифмическими координатами данной группы.

Полезно отметить также следующее. При доказательстве теоремы мы
предполагали, что квантовая скорость хорошо определена, то есть
что производная вектора по параметру существует и принадлежит
гильбертову пространству. Однако, если генератор является
неограниченным оператором, для некоторых векторов состояния это
может оказаться и не так. Тем не менее, утверждение теоремы
оказывается верным и в этом случае, поскольку в этом случае
величина $\Delta_\rangle A$ оказывается бесконечной. Поэтому при
ненулевом значении параметра в левой части неравенства оказывается
бесконечность (которая, конечно, больше постоянной Планка).

Таким образом, содержание теоремы оказывается верным для {\it
любого} вектора состояния $\rangle$ и {\it любого}
(самосопряжённого) генератора $A$.

\section*{Подгруппы группы Пуанкаре}

Приведённая в конце предыдущего раздела теорема и является самым
общим математическим выражением принципа определённости. При этом,
в зависимости от конкретной задачи, рассматриваемая унитарная
группа может являться представлением той или иной физической
группы симметрии. Наиболее важный пример при этом представляет
собой группа Пуанкаре, общее соотношение определённости для
которой (\ref{CertPoin}) уже приводилось выше. Эта группа имеет
некоторые физически важные подгруппы, которые настолько
практически важны, что мы обсудим их здесь более подробно.


\paragraph{Пространственные трансляции.}
Генераторами трёхпараметрической группы пространственных
трансляций, как известно, являются компоненты векторного оператора
импульса $P_1$, $P_2$ и $P_3$. Соотношение определённости,
соответственно, записывается как
\begin{equation}
 \Delta_\rangle (\,
   \delta x_1\,P_1\, +\,
   \delta x_2\,P_2\, +\,
   \delta x_3\,P_3\,
 ) \geqslant \hbar  \ .
 \label{CertTrans}
\end{equation}
Здесь числа $\delta x_1$, $\delta x_2$ и $\delta x_3$ составляют
компоненты вектора, который задаёт пространственный сдвиг.

Если положить теперь $\delta x_2=\delta x_3=0$, а также опустить
для краткости индекс, то неравенство (\ref{CertTrans}) можно
записать в виде:
\[
 |\delta x|\, \Delta_\rangle P \geqslant \hbar \ .
\]

Допустим теперь, что у рассматриваемой квантовой системы нам
удалось найти некоторую наблюдаемую $X$, которую можно в том или
ином смысле считать её \rql оператором координаты\rqr. Это
предположение принимается как само собой разумеющееся в
нерелятивистской квантовой механике. В релятивистской же квантовой
теории никакого естественного понятия координаты нет.\footnote{Это
касается в том числе и дираковского электрона. Суперпозиция
положительно- и отрицательно-частотного решений уравнения Дирака
не имеет никакого физического смысла, а потому наблюдаемой,
описываемой \rql оператором умножения на переменную $x$\rqr{} не
существует.} Можно, например, формально составить из генераторов
группы Пуанкаре векторный оператор, компонентам которого в
классической механике соответствуют координаты центра масс. При
этом следует иметь в виду, что он будет обладать довольно
необычными свойствами. Например, его компоненты не будут
коммутировать друг с другом.

Будем полагать, что $X$~--- самосопряжённый оператор с непрерывным
спектром, $X=X^*$. Обозначим $\Omega_{(a,b)}$ его спектральный
проектор\footnote{Грубо говоря, спектральный проектор~--- это
оператор, зануляющий волновую функцию в $X$-представлении вне
данного интервала.} для произвольного вещественного интервала
$(a,b)$.

Предположим также, что $X$, будучи координатой, ведёт себя обычным
образом под действием трансляций, а именно, что для любых $a$, $b$
и $\delta x$ имеет место равенство:
\[
 e^{+\,i\,\delta x\,P/\hbar} \,
  \Omega_{(a+\delta x,b+\delta x)} \,
    e^{-\,i\,\delta x\,P/\hbar} \,=\,
 \Omega_{(a,b)} \ ,
\]
то есть $P$ является \rql генератором спектральных сдвигов\rqr{}
для $X$.

Пусть теперь система находится в состоянии $\rangle$,
$\langle\,|\,\rangle=1$.

Величина $\langle\,\Omega_{(a,b)}\,\rangle$, очевидно, определяет
вероятность найти систему внутри интервала $(a,b)$. Определим
такие $l$ и $r$, что
\[
 \textstyle
 \langle\,\Omega_{(-\infty,l)}\,\rangle
 \,=\,
 \langle\,\Omega_{(r,+\infty)}\,\rangle
 \,=\,
 \frac{1-\sin1}2
 \,\approx\, 0,07926\dots
\]
Легко видеть, что $l$ и $r$ существуют\footnote{Но, вообще говоря,
не единственны. Чтобы устранить эту многозначность, удобно
выбирать $l$ максимальным из возможных, а $r$~--- минимальным.
Тогда расстояние $r-l$ будет минимальным.}. При этом величину
$\delta_\rangle X = r-l$ вполне естественно можно назвать \rql
неопределённостью\rqr{} координаты $X$.

\mytheorem{}(принцип неопределённости)
 {\it Имеет место неравенство:
\begin{equation}
 \delta_\rangle X \Delta_\rangle P \geqslant \hbar \ .
 \label{Uncertainty}
\end{equation}}

Для доказательства этой теоремы воспользуемся (двукратно)
неравенством треугольника для квантового угла:
\[
 \angle\,\bigl(\
  \rangle
  \ , \
  e^{-\,i\,\delta_\rangle X\,P/\hbar}\,\rangle
 \ \bigr)
 \ \geqslant\
 \angle\,\bigl(\
  \Omega_{(-\infty,r)}\,\rangle
  \ , \
  \Omega_{(r,+\infty)}\,e^{-\,i\,\delta_\rangle X\,P/\hbar}\,\rangle
 \ \bigr)
 \,-\,
 \qquad\qquad\qquad\qquad\qquad\qquad
\]
\[
 \qquad\qquad\qquad\qquad\qquad\qquad
 \,-\,
 \angle\,\bigl(\
  \rangle
  \ , \
  \Omega_{(-\infty,r)}\,\rangle
 \ \bigr)
 \,-\,
 \angle\,\bigl(\
  e^{-\,i\,\delta_\rangle X\,P/\hbar}\,\rangle
  \ , \
  \Omega_{(r,+\infty)}\,e^{-\,i\,\delta_\rangle X\,P/\hbar}\,\rangle
 \ \bigr)
 \,=\,
\]
\[
 \textstyle
 \,=\,
 \frac\pi2
 \,-\,
 \arcsin\sqrt{\frac{1-\sin1}2}
 \,-\,
 \arcsin\sqrt{\frac{1-\sin1}2}
 \,=\,
 \frac\pi2
 \,-\,
 \left(\frac\pi4-\frac12\right)
 \,-\,
 \left(\frac\pi4-\frac12\right)
 \,=\,
 1 \ .
\]
Но это означает, что под действием
 $e^{-\,i\,\delta_\rangle X\,P/\hbar}$
вектор $\rangle$ меняется существенно.

Применяя принцип {\it определённости}, непосредственно получаем
(\ref{Uncertainty}). $\blacksquare$

Таким образом, принцип определённости позволяет получить
альтернативное неравенство (\ref{Uncertainty}), выражающее принцип
неопределённости. Это неравенство не эквивалентно неравенству
Кеннарда ({\ref{KennardRel}).

С формально-математической точки зрения, ни одно из неравенств
({\ref{KennardRel}) и (\ref{Uncertainty}) не является сильнее
другого.

Однако, нетрудно видеть, что отношение величин $\Delta_\rangle X$
и $\delta_\rangle X$ всегда подчинено неравенству:
\[
 \frac{2\,\Delta_\rangle X}{\delta_\rangle X} \,\geqslant\,
 \sqrt{1-\sin1} \,\approx\, 0,398\dots
\]
Следовательно, если ослабить неравенство Кеннарда
({\ref{KennardRel}) дополнительным множителем $0,398$ (что обычно
непринципиально для качественных оценок), то оно становится слабее
(\ref{Uncertainty}).

С другой стороны, отношение $\Delta_\rangle X/\delta_\rangle X$
может быть сколь угодно большим, и даже бесконечно большим, если
волновой пакет плохо локализован. В такой ситуации неравенство
Кеннарда ({\ref{KennardRel}), в отличие от неравенства
(\ref{Uncertainty}), вообще не даёт никакой оценки для
неопределённости импульса.

Таким образом, неравенство (\ref{Uncertainty}) является более
содержательным, чем ({\ref{KennardRel}), для качественных оценок.


\paragraph{Пространственные вращения.}
Генераторами трёхпараметрической группы пространственных вращений
являются компоненты векторного оператора момента импульса $J_1$,
$J_2$ и $J_3$. Соотношение определённости, соответственно,
записывается как
\[
 \Delta_\rangle (\,
   \delta\varphi_1\,J_1\, +\,
   \delta\varphi_2\,J_2\, +\,
   \delta\varphi_3\,J_3\,
 ) \geqslant \hbar  \ .
\]
Логарифмические координаты имеют здесь простую геометрическую
интерпретацию: поворот, описываемый координатами
$\delta\varphi_1$, $\delta\varphi_2$ и $\delta\varphi_3$ можно
выполнить путём поворота вокруг вектора с этими координатами на
угол, равный длине этого вектора.

Полагая $\delta\varphi_2=\delta\varphi_3=0$ и опуская индекс,
имеем:
\[
 |\delta\varphi|\, \Delta_\rangle J \geqslant \hbar \ .
\]

Попробуем теперь получить соотношение неопределённостей по той же
схеме, что и в случае группы пространственных трансляций.

Прежде всего необходимо заметить, что соотношения {\it
неопределённости} в форме \rql произведение неопределённости угла
на неопределённость момента импульса больше либо равно постоянной
Планка\rqr{} предложить заведомо невозможно, поскольку в
состоянии, когда неопределённость момента импульса обращается в
ноль (собственное состояние), \rql неопределённость угла\rqr{},
как бы она ни определялась, не должна превосходить $2\pi$.

Далее, здесь мы сталкиваемся с той трудностью, что \rql
хороший\rqr{} самосопряжённый оператор угла ввести невозможно.
Операторов угла оказывается бесконечно много (если определён хотя
бы один), но как только мы фиксируем выбор одного из них
каким-нибудь условием (например, чтобы его спектр лежал в
интервале от $0$ до $2\pi$), он оказывается довольно сложно
преобразующимся под действием вращений, и анализ формул сильно
затрудняется.

Поэтому мы не будем здесь вообще уточнять выбор одного из
бесконечного множества операторов угла, а будем подразумевать, что
речь идёт сразу обо всём бесконечном классе. При этом для этого
класса мы будем использовать символическое обозначение $\Phi$.

Такой подход не порождает трудностей, поскольку при выводе
соотношения неопределённости нам фактически приходится оперировать
со спектральной мерой (множеством спектральных проекторов). А эта
спектральная мера у всех операторов угла, в сущности, одна и та
же. Просто аргументом этой меры фактически выступает не отрезок
вещественной прямой, как в случае с оператором координаты, а
сегмент окружности. Вещественная координата же на окружности
должна пониматься с точностью до слагаемого $2\pi$. Таким образом,
спектральная мера удовлетворяет соотношению:
\[
 \Omega_{(a+2\pi,b+2\pi)} = \Omega_{(a,b)} \ .
\]

Предположим также, что оператор $J$ момента импульса является \rql
генератором спектральных вращений\rqr{} для $\Phi$:
\[
 e^{+\,i\,\delta\varphi\,J/\hbar} \,
  \Omega_{(a+\delta\varphi,b+\delta\varphi)} \,
    e^{-\,i\,\delta\varphi\,J/\hbar} \,=\,
 \Omega_{(a,b)} \ .
\]

Величина $\langle\,\Omega_{(a,b)}\,\rangle$, очевидно, определяет
вероятность найти систему внутри углового сегмента $(a,b)$. Пусть
теперь имеется такой небольшой угловой сегмент $(l,r)$, в котором
в основном сосредоточена вероятность обнаружить систему. А именно,
предположим, что вероятность обнаружить систему в дополнительном
сегменте равна:
\[
 \textstyle
 \langle\,\Omega_{(r,l+2\pi)}\,\rangle
 \,=\,
 \frac{1-\sin1}2
 \,\approx\, 0,07926\dots
\]
Если $r$ и $l$ выбраны так, что величина
 $\delta_\rangle \Phi = r-l$ минимальна, то
$\delta_\rangle \Phi$ можно при этом назвать \rql
не\-опре\-де\-лён\-ностью\rqr{} угла $\Phi$.

\mytheorem{} (принцип неопределённости для угла и момента
импульса)
 {\it Имеет место неравенство:
\begin{equation}
 \delta_\rangle\Phi \ \geqslant\
 \min\ (\ \hbar\,/\,\Delta_\rangle J\ ;\ \pi\ ) \ .
 \label{UncertRot}
\end{equation}}

Рассмотрим случай, когда $\delta_\rangle\Phi\,\leqslant\,\pi$. При
этом повёрнутый угловой сегмент $(l,r)$ не будет перекрываться с
неповёрнутым, и можно применить тот же метод оценки квантового
угла, что и в случае трансляций.

Итак, воспользуемся (двукратно) неравенством треугольника для
квантового угла:
\[
 \angle\,\bigl(\
  \rangle
  \ , \
  e^{-\,i\,\delta_\rangle\Phi\,J/\hbar}\,\rangle
 \ \bigr)
 \ \geqslant\
 \angle\,\bigl(\
  \Omega_{(l,r)}\,\rangle
  \ , \
  \Omega_{(r,r+\delta_\rangle\Phi)}\,e^{-\,i\,\delta_\rangle\Phi\,J/\hbar}\,\rangle
 \ \bigr)
 \,-\,
 \qquad\qquad\qquad\qquad\qquad\qquad
\]
\[
 \qquad\qquad\qquad\qquad\qquad\qquad
 \,-\,
 \angle\,\bigl(\
  \rangle
  \ , \
  \Omega_{(l,r)}\,\rangle
 \ \bigr)
 \,-\,
 \angle\,\bigl(\
  e^{-\,i\,\delta_\rangle\Phi\,J/\hbar}\,\rangle
  \ , \
  \Omega_{(r,r+\delta_\rangle\Phi)}\,e^{-\,i\,\delta_\rangle\Phi\,J/\hbar}\,\rangle
 \ \bigr)
 \,=\,
\]
\[
 \textstyle
 \,=\,
 \frac\pi2
 \,-\,
 \arcsin\sqrt{\frac{1-\sin1}2}
 \,-\,
 \arcsin\sqrt{\frac{1-\sin1}2}
 \,=\,
 \frac\pi2
 \,-\,
 \left(\frac\pi4-\frac12\right)
 \,-\,
 \left(\frac\pi4-\frac12\right)
 \,=\,
 1 \ .
\]
Но это означает, что под действием
 $e^{-\,i\,\delta_\rangle\Phi\,J/\hbar}$
вектор $\rangle$ меняется существенно.

Применяя принцип {\it определённости}, непосредственно получаем
(\ref{UncertRot}). $\blacksquare$

Как видим, принцип определённости позволяет легко получить и
простое соотношение {\it неопределённости} для угла и момента
импульса.

Следует отметить, что попытка \cite{Judge1964} получить
соотношение неопределённости для угла и момента импульса по
аналогии с неравенством Кеннарда ({\ref{KennardRel}) приводит к
довольно сложным формулам. А именно, если зафиксировать
самосопряжённый оператор угла $\Phi$ условием, что его спектр
является интервалом $[-\pi,\pi]$, и определить \rql
неопределённость угла\rqr{} по формуле:
\[
 \Delta'_\rangle\Phi =
  \min_{\delta\varphi}
  \sqrt{\langle\,
   e^{+\,i\,\delta\varphi\,J/\hbar} \,
   \Phi^2 \,
   e^{-\,i\,\delta\varphi\,J/\hbar}\,\rangle} \ ,
\]
то имеет место соотношение неопределённости:
\begin{equation}
 \Delta'_\rangle\Phi \, \Delta_\rangle J \,\geqslant\,
 \frac\hbar2
 \,
 \left[\,
  1\, -\, \frac3{\pi^2}\, (\Delta'_\rangle\Phi)^2
 \,\right]
 \ .
 \label{UncJudge}
\end{equation}
Для исследования квазиклассического предела неравенство
(\ref{UncertRot}) оказывается более удобным, чем (\ref{UncJudge}),
т.~к. позволяет накладывать более мягкие условия на локализацию
волнового пакета.


\paragraph{Временные сдвиги.}
Группа Пуанкаре включает в качестве подгруппы однопараметрическую
группу сдвигов во времени. Естественно ожидать, что принцип
определённости имеет соответствующее математическое выражение и в
этом случае. И это действительно так.

Но наше обсуждение здесь осложняется тем обстоятельством, что вся
учебная литература и по классической и по квантовой механике
ориентируется исключительно на решение задач динамики. При этом
понятие динамического состояния обязательно относят к некоторому
моменту времени. Такой подход полностью скрывает релятивистскую
инвариантность. При этом понятие группы временных сдвигов
становится фактически бессодержательным.

Полностью последовательными в этом плане являются инвариантный
гамильтонов формализм (в случае классической механики) и
релятивистское каноническое квантование (РКК) (в случае
релятивистской квантовой теории поля). При этом в теорию
естественным образом включаются не только временные сдвиги, но и
лоренцевы бусты. Однако, поскольку с таким подходом в настоящее
время большинство физиков знакомы плохо, мы здесь обсудим
промежуточный подход, который, в сущности, основан на
представлении Гейзенберга. (Лоренцевы бусты при этом мы в этом
обзоре вообще рассматривать не будем.)

Обычно под динамическим состоянием $S_t$ в некоторый момент
времени $t$ понимают набор величин, которые можно измерить в
момент времени $t$ некоторым набором приборов, и которые задают
всю последующую эволюцию системы (то есть аналогичные наборы
величин в последующие моменты времени). В классической механике
динамическое состояние обычно описывается либо набором координат и
скоростей (лагранжев формализм), либо набором координат и
импульсов (гамильтонов формализм). То есть динамическое состояние
в заданный момент времени определяется точкой на некотором
многообразии, называемом фазовым пространством. В квантовой
механике динамическое состояние описывается вектором в
гильбертовом пространстве.

Когда встаёт вопрос об описании действия группы пространственных
движений на пространстве динамических состояний в некоторый момент
времени, обычно рассуждают так. Пусть набор физических приборов
$K$, измеряющих полный набор динамических переменных
рассматриваемой системы, под действием некоторого движения $G$
перешёл в $K'$, $G:K\to K'$. И пусть для некоторого динамического
состояния системы $S$ имеется динамическое состояние $S'$, такое
что набор приборов $K'$, измеряющий параметры состояния $S'$,
выдаёт те же результаты, что и $K$, измеряющий параметры состояния
$S$. Тогда говорят, что под действием преобразования $G$ состояние
$S$ переходит в $S'$. Символически это можно записать так:
\[
 G:K\to K' \ ,\ K(S)=K'(S')
 \qquad\Longrightarrow\qquad
 G:S\to S' \ .
\]
Именно таким путём вводится действие пространственных сдвигов и
вращений, описанное выше.

Казалось бы, это же определение естественно можно распространить и
на группу временных сдвигов:
\[
 G:K_{t_1}\to K_{t_2} \ ,\ K_{t_1}(S_{t_1})=K_{t_2}(S_{t_2})
 \qquad\Longrightarrow\qquad
 G:S_{t_1}\to S_{t_2} \ .
\]
Но от этого определения оказывается мало пользы, поскольку оно
оказывается чисто формальным и никак не связанным с динамикой.

Приведённое выше определение динамического состояния мы немного
усовершенствуем. Будем считать, что если некоторое динамическое
состояние $S_{t_1}$ в ходе эволюции переходит в $S_{t_2}$, то
$S_{t_1}$ и $S_{t_2}$ описывают одно и то же динамическое
состояние. То есть динамическое состояние~--- это просто некоторый
абстрактный набор величин (не обязательно связанный с измерениями
в конкретный момент времени), определяющий эволюцию системы. Когда
речь идёт о квантовых системах, можно считать, что динамическое
состояние просто описывается гейзенберговским вектором состояния.

При этом, если система замкнута, то её гамильтониан не зависит от
времени, а также коммутирует с пространственными движениями.
Поэтому описанная в предыдущих пунктах теория пространственных
движений переносится и на этот случай. Кроме того, сюда
присоединяется ещё и группа временных сдвигов, задаваемая
семейством операторов:
\[
 U(\delta t) = e^{-\,i\,\delta t\,(-H)/\hbar} \ .
\]
Следует обратить внимание, что генератором этой группы является
гамильтониан со знаком \rql минус\rqr{}. Оператор, описывающий
действие рассматриваемой группы, является обратным к оператору,
описывающему эволюцию в уравнении Шрёдингера. Можно сказать, что
здесь группа действует активно, в то время как в представлении
Шрёдингера она действует пассивно.

Соотношение определённости для группы временных сдвигов принимает
вид:
\begin{equation}
 |\delta t|\, \Delta_\rangle H \geqslant \hbar \ .
 \label{CertTH}
\end{equation}
Если посмотреть на это соотношение с точки зрения представления
Шрёдингера (при этом естественная связь с группой Пуанкаре
утрачивается), то это соотношение как раз оказывается соотношением
Мандельштама-Тамма \cite{MandTamm1945} для скорости квантовой
эволюции.

Таким образом, соотношение Мандельштама-Тамма для случая замкнутых
систем является следствием принципа определённости.

Обсудим теперь незамкнутые системы. Соотношение Мандельштама-Тамма
для скорости квантовой эволюции (получаемое в представлении
Шрёдингера) естественным образом обобщается на этот случай.
Обозначим гамильтониан в этом случае как $H^{\mathrm{full}}$. При
этом, поскольку величина $\Delta_{|t\rangle} H^{\mathrm{full}}$,
вообще говоря, зависит от времени, её нужно просто усреднить:
\begin{equation}
 |\tau|\, \overline{\Delta_{|t\rangle} H^{\mathrm{full}}}
 \geqslant \hbar \ .
 \label{MTFull}
\end{equation}
Здесь $\tau$~--- промежуток времени, в течение которого происходит
эволюция (он не имеет смысла логарифмической координаты на
какой-либо группе).

Принцип определённости в этой ситуации оказывается формально
неприменим. Однако, на практике, когда рассматриваются незамкнутые
квантовые системы, их гамильтониан обычно представим в виде суммы
двух слагаемых, собственного гамильтониана \rql свободной\rqr{}
системы и гамильтониана взаимодействия \rql с внешним классическим
полем\rqr{}:
\[
 H^{\mathrm{full}} = H + H^{\mathrm{int}} \ .
\]
При этом гамильтониан взаимодействия $H^{\mathrm{int}}$ обычно
можно \rql выключать\rqr{}.

В этом случае естественно связать понятие динамического состояния
системы с её эволюцией при выключенном гамильтониане
взаимодействия (хотя рассматриваться оно может в конкретный момент
времени, без выключения гамильтониана взаимодействия).
Соответственно, соотношение определённости (\ref{CertTH})
сохраняет свой прежний вид. Смысл его становится несколько более
абстрактным, без какой либо прямой связи с соотношением
Мандельштама-Тамма (\ref{MTFull}).

В заключение отметим, что Мандельштам и Тамм вывели своё
соотношение путём прямой оценки скалярного произведения для
начального и конечного векторов состояния. Теорию метрики
Фубини-Штуди они при этом не использовали. Такой вывод оказывается
существенно короче. Но при этом геометрическая природа результата
оказывается в значительной степени скрытой. Это дополнительно
усложняет интерпретацию (помимо более серьёзного недостатка~---
интерпретации, основанной на уравнении Шрёдингера).

Вывод соотношения Мандельштама-Тамма (в рамках уравнения
Шрёдингера), основанный на метрике Фу\-би\-ни-Штуди, был дан
позднее Ананданом и Аароновым \cite{AnanAhar1990}. При этом ссылки
на Мандельштама и Тамма не давалось. Поэтому в западной литературе
это соотношение часто ошибочно называют соотношением
Анандана-Ааронова.

\section*{Благодарности}

В заключение хочу поблагодарить А.~Клейна, С.~Лоренца,
Т.~А.~Болохова, А.~В.~Осипова, Е.~В.~Аксёнову, А.~Ю.~Тощевикову,
М.~Гатти и А.~К.~Пэти за полезные дискуссии, литературные указания
и помощь в добывании литературы.


\section*{}

\tcptab
